\documentclass[aps,showpacs,nofootinbib,superscriptaddress]{revtex4}
\usepackage{epsf}
\usepackage{graphicx}
\usepackage{amsmath}
\usepackage[utf8]{inputenc}
\usepackage{times}
\usepackage[T1]{fontenc}
\def\tstrut{\vrule height2.5ex depth0pt width0pt} 
\def\slashchar#1{\setbox0=\hbox{$#1$}
   \dimen0=\wd0 \setbox1=\hbox{/} \dimen1=\wd1
   \ifdim\dimen0>\dimen1 \rlap{\hbox to \dimen0{\hfil/\hfil}} #1
   \else  \rlap{\hbox to \dimen1{\hfil$#1$\hfil}} / \fi}

\begin{document}

\title{Theoretical study of neutrino-induced coherent pion production
off nuclei at T2K and MiniBooNE energies} 

\author{J.~E.~Amaro}
\affiliation{Departamento de F{\'\i}sica At\'omica, Molecular y
Nuclear,\\ Universidad de Granada, E-18071 Granada, Spain} 
\author{E. Hern\'andez} 
\affiliation{Grupo de F\'\i sica Nuclear, Departamento
de F\'\i sica Fundamental e IUFFyM,\\ Facultad de Ciencias, E-37008
Salamanca, Spain.}  
\author{J.~Nieves} 
\affiliation{ Instituto de F\'\i sica Corpuscular (IFIC), Centro Mixto
  CSIC-Universidad de Valencia, Institutos de Investigaci\'on de
  Paterna, Aptd. 22085, E-46071 Valencia, Spain} 
\author{M.~Valverde} 
\affiliation{Departamento de F{\'\i}sica At\'omica,
Molecular y Nuclear,\\ Universidad de Granada, E-18071 Granada, Spain}
\affiliation{Department of Subatomic and Radiation Physics, \\
Ghent University, Proeftuinstraat 86, B-9000 Gent, Belgium}

\pacs{25.30.Pt,13.15.+g,12.15.-y,12.39.Fe}

\today

\begin{abstract}
  We have developed a model for neutrino-induced coherent pion
  production off nuclei in the energy regime of interest for present
  and forthcoming neutrino oscillation experiments. It is based on a
  microscopic model for pion production off the nucleon that, besides
  the dominant $\Delta$ pole contribution, takes into account the
  effect of background terms required by chiral symmetry. Moreover,
  the model uses a reduced nucleon-to-$\Delta$ resonance axial
  coupling, which leads to coherent pion production cross sections
  around a factor two smaller than most of the previous theoretical
  estimates. In the coherent production, the main nuclear effects,
  namely medium corrections on the $\Delta$ propagator and the final
  pion distortion, are included. We have improved on previous similar
  models by taking into account the nucleon motion and employing a
  more sophisticated optical potential.  As found in previous
  calculations the modification of the $\Delta$ self-energy inside the
  nuclear medium strongly reduces the cross section, while the final
  pion distortion mainly shifts the peak position to lower pion
  energies. The angular distribution profiles are not much affected by
  nuclear effects.  Nucleon motion increases the cross section by
  $\sim 15 $\% at neutrino energies of 650\, MeV, while Coulomb
  effects on charged pions are estimated to be small. Finally, we
  discuss at length the deficiencies of the Rein--Sehgal pion coherent
  production model for neutrino energies below 2 GeV, and in
  particular for the MiniBooNE and T2K experiments. We also predict flux
  averaged cross sections for these two latter experiments and K2K.
\end{abstract}

\maketitle

\section{Introduction}

Neutrino-induced one-pion production off nucleons and nuclei in the
intermediate energy region is a source of relevant data on hadronic
structure. Pions are mainly produced through resonance excitation and
these reactions can be used to extract information on
nucleon-to-resonance axial transition form factors. Besides, a proper
understanding of these processes is very important in the analysis of
neutrino oscillation experiments. For instance, $\pi^0$ production by
neutral currents (NC) is the most important $\nu_\mu$-induced
background to experiments that measure $\nu_\mu\to\nu_e$ oscillations
in the neutrino energy range around
$1$~GeV~\cite{AguilarArevalo:2007it}. This is because NC $\pi^0$
events can mimic $\nu_e$ signal events when, for example, one of the
two photons associated with the $\pi^0\to \gamma\gamma$ decay is not
detected. This can happen when a photon exits the detector before
showering or does not have enough energy to initiate a shower.
Similarly, $\pi^+$ production by charged currents (CC) is an important
source of background in $\nu_\mu\to \nu_x$ disappearance
searches~\cite{Hiraide:2006zq}.

In reactions on nuclei, pions can be produced incoherently or
coherently. In the latter case the nucleus remains in its ground
state. Coherent reactions are controlled by the nucleus form factor
and are more forward peaked than incoherent ones.  CC coherent pion
production has been studied at higher energies in a number of
experiments~\cite{Marage:1986cy,Grabosch:1985mt,Allport:1988cq,Aderholz:1988cs,Vilain:1993sf,Willocq:1992fv}.
The results could be satisfactorily explained by the Rein--Sehgal
model~\cite{Rein:1982pf} which is based on the partially conserved
axial current (PCAC) hypothesis~\cite{Adler:1964yx}.  The K2K
Collaboration has recently conducted a search for CC coherent pion
production induced by muon neutrinos with a mean beam energy of $1.3$
GeV~\cite{Hasegawa:2005td}. Contrary to expectations, they found no
evidence for CC coherent pion production, setting an upper limit of
0.60\% for the coherent to total CC pion production ratio.  The data
show a deficit of forward muons in the kinematical region where a
sizable coherent production is expected. An attempt to explain this
deficit has been done by Rein and Sehgal in Ref.~\cite{Rein:2006di} by
including in their model the usually neglected finite muon mass
effect~\cite{Adler:2005ada,Berger:2007rq}. In this way they find a
$25\%$ suppression caused by the destructive interference between the
axial vector and pseudoscalar (pion-pole) amplitudes, reducing in this
way the discrepancy between theory and experiment, though it still
persists.  This correction affects only CC processes and its relevance
is reduced as the neutrino energy increases~\cite{Rein:2006di}.  The
negative K2K results are consistent with a very recent search
performed by the SciBooNE Collaboration~\cite{Hi08}. 

NC coherent pion production was observed by the Aachen-Padova
group~\cite{Faissner:1983ng} on a $^{27}$Al target with both the muon
neutrino and antineutrino CERN PS beam with average energy of $2$
GeV. Positive evidence was also seen by the PS-Gargamelle neutrino and
antineutrino Freon experiments~\cite{Isiksal:1984vh}.  Very recently, the
MiniBooNE Collaboration announced the first observation of NC coherent
$\pi^0$ production below $2$ GeV~\cite{AguilarArevalo:2008xs}. When
integrated over the MiniBooNE flux, they find a ratio of coherent plus
diffractive production over all exclusive NC $\pi^0$ production given
by $19.5\pm1.1(stat.)\pm2.5(sys.)$\% for a mineral oil target
(CH$_2$).  By using Monte Carlo they estimate the coherent rate for a
pure $^{12}$C target to be $20.3\pm2.8(stat.)$\%. 

On the theoretical side the Rein--Sehgal model~\cite{Rein:1982pf}
mentioned above assumes that coherent pion production is dominated by
the divergence of the axial
current~\cite{Nachtmann:1970yv,Bell:1964eu,Pais:1974kd,Lackner:1979ax}
and can thus be related to the pion-nucleus coherent scattering via
PCAC. Extrapolation to non-forward angles is done by including a
propagator term $(1+Q^2/m_A^2)^{-2}$ with $m_A\approx 1$\,GeV. The
effects on the model of considering a finite muon mass was recently
analysed in Refs.~\cite{Rein:2006di,Berger:2007rq} and, as stated
above, they give rise to a $25\% $ 
reduction of the CC coherent pion
production by muon neutrinos at low neutrino energies. However, one
should note that the Rein--Sehgal model does not account for
nuclear pion absorption, since it does not consider two body
mechanisms which are those responsible for the absorption of the
outgoing pions, and it does not correctly treat
quasielastic collisions either. Besides,  the corrections to 
the outgoing pion angular dependence predicted by the model become quite
important for the low neutrino energies relevant in MiniBooNE and T2K
experiments, as we will show in Subsect.~\ref{sec:rs}.

The PCAC approach was also used in the models of
Refs.~\cite{Belkov:1986hn,Kopeliovich:2004px,Gershtein:1980vd,Komachenko:1983jv,Paschos:2005km}.
In Ref.\cite{Paschos:2005km} the authors take into account the muon
mass effect and include a small non-PCAC transverse current
contribution. In all cases the distortion of the final pion was
included. There are other approaches that do not rely on PCAC.  In
Ref.~\cite{Kim:1996az} coherent pions are produced by virtual
$\Delta$-h excitations in the nucleus. The model includes the
modifications of the nucleon and $\Delta$ propagators in the medium,
evaluated in a relativistic mean field approximation, but no final
pion distortion was taken into account.  Kelkar {\it et
al.}~\cite{Kelkar:1996iv} improve on the above calculation by doing a
more sophisticated evaluation of the $\Delta$ self-energy in the
medium and treating the final pion distortion in a realistic way by
solving the Klein-Gordon (KG) equation for a pion-nucleus optical
potential.  The model of Ref.~\cite{Singh:2006bm} uses similar medium
corrections and improves on the description of the elementary
reaction.  On the other hand the final pion distortion is treated in
the eikonal approximation which is known to fail at low pion energies.
In Refs.~\cite{AlvarezRuso:2007tt,AlvarezRuso:2007it} the authors
follow the Kelkar {\it et al.}  calculation in the treatment of the
final pion distortion while using, as in
Ref.~\cite{Singh:2006bm}, a more complete
and fully relativistic elementary amplitude.

In a recent publication we have developed a model for CC and NC
neutrino- and antineutrino-induced pion production off the nucleon in
the intermediate energy region~\cite{Hernandez:2007qq}, which
represents the natural extension of that developed in
Ref.~\cite{Gil:1997bm} for the electron analogue $e N \to e' N' \pi$
reaction. Most previous studies~\cite{Adler:1968tw,Llewellyn
Smith:1971zm,Schreiner:1973ka,AlvarezRuso:1997jr,AlvarezRuso:1998hi,Lalakulich:2005cs,Leitner:2006ww,Paschos:2003qr,Lalakulich:2006sw}
of these processes considered only the dominant $\Delta$ pole mechanism
in which the neutrino excites a $\Delta(1232)$ resonance that
subsequently decays into $N\pi$. In our model we have also included
background terms required by chiral symmetry (See
Fig.~\ref{fig:diagramas} below). Some background terms were considered
before in the works of
Refs.~\cite{Fogli:1979cz,Fogli:1979qj,Sato:2003rq} although none of
these models was consistent with chiral counting.  In
Ref.~\cite{Hernandez:2007qq}, we found that background terms produced
significant effects in all channels. As a result we had to re adjust
the strength of the dominant $\Delta$ pole contribution.  The least known
ingredients of the model are the axial nucleon-to-$\Delta$ transition
form factors, of which $C_5^A$ gives the largest contribution. This
strongly suggested the readjustment of that form factor to the experimental
data, which we did by fitting the flux-averaged $\nu_\mu p\to \mu^- p
\pi^+$ ANL $q^2$-differential cross section for pion-nucleon invariant
masses $W < 1.4$ GeV~\cite{Barish:1978pj,Radecky:1981fn}. Our full
model, thus obtained, lead to an overall better description of the
data for different CC and NC, neutrino- and antineutrino-induced,
one-pion production reactions off the nucleon. This reduction of the
$C_5^A(0)$ value is consistent with recent results in lattice QCD
\cite{Alexandrou:2006mc}, quark model \cite{BarquillaCano:2007yk} and
phenomenological studies \cite{Graczyk:2007bc}.
 
Here, we shall apply our model to evaluate CC and NC coherent pion
production in nuclei, including the chiral background terms in the
elementary amplitude. We follow a scheme similar to that advocated in
Refs.~\cite{AlvarezRuso:2007tt,AlvarezRuso:2007it}\footnote{ We use a
  realistic description of the $\Delta$ self-energy in the medium and
  treat quantum mechanically the final pion distortion.}, but
improving the results of these latter references by properly taking into
account the  motion of the nucleons and correcting for
some numerical inaccuracies that affected the calculations of these
two references~\cite{luis}. Our model should work better close to
threshold and hence we will concentrate in the neutrino energy range
of MiniBooNE and the future T2K experiment where the neutrino peak energy is
expected to be around $0.6\div 0.7$ GeV~\cite{Kato:2007zzc}. This work
is organized as follows: in section~\ref{sec:cc} we discuss our model
for the evaluation of CC coherent pion production, including the most
relevant aspects of medium corrections for the $\Delta$ and the
evaluation of the final pion distortion. In section~\ref{sec:nc} we
find the corresponding expressions for the NC coherent pion production
case. Finally, in section~\ref{sec:results} we present and discuss our
results.

\section{CC Neutrino and Antineutrino induced reactions}

\label{sec:cc}

We will focus on the coherent CC pion production reaction induced by
neutrinos,
\begin{equation}
  \nu_l (k) +\, A_Z|_{gs}(p_A)  \to l^- (k^\prime) +
  A_Z|_{gs}(p^\prime_A) +\, \pi^+(k_\pi) 
\label{eq:reac}
\end{equation}
The process consists of a weak pion ($\pi^+$) production followed by the strong
distortion of the pion in its way out of the nucleus. In the coherent
production the nucleus is left in its ground state by contrast with
the incoherent production where the nucleus is either broken or left
in some excited state.

The unpolarized differential cross section, with respect to the
outgoing lepton and pion kinematical variables, is given in the
Laboratory (LAB) frame by
\begin{equation}
\label{eq:sec}
\frac{d^{\,5}\sigma_{\nu_l
    l}}{d\Omega(\hat{k^\prime})dE^\prime d\Omega(\hat{k}_\pi) } =
    \frac{|\vec{k}^\prime|}{|\vec{k}~|}\frac{G^2}{4\pi^2}
      {L}_{\mu\sigma}^{(\nu)}\, {W}^{\mu\sigma}_{{\rm CC}\pi^+} 
\end{equation}
with $\vec{k}$ and $\vec{k}^\prime~$ the LAB lepton momenta,
$E^{\prime} = (\vec{k}^{\prime\, 2} + m_l^2 )^{1/2}$ and $m_l$ the
energy and the mass of the outgoing lepton, $G=1.1664\times 10^{-11}$
MeV$^{-2}$ the Fermi constant, $\vec{k}_\pi$ and $E_\pi=
(\vec{k}^{2}_\pi + m_\pi^2 )^{1/2}$ the LAB momentum and energy of the
outgoing pion, and ${ L}$ and ${ W}$ the leptonic and hadronic
tensors, respectively. The leptonic tensor is given by (in our
convention, we take $\epsilon_{0123}= +1$ and the metric
$g^{\mu\nu}=(+,-,-,-)$):
\begin{eqnarray}
{ L}_{\mu\sigma}^{(\nu)}
=
 k^\prime_\mu k_\sigma +k^\prime_\sigma k_\mu
- g_{\mu\sigma} k\cdot k^\prime + {\rm i}
\epsilon_{\mu\sigma\alpha\beta}k^{\prime\alpha}k^\beta \label{eq:lep}
\end{eqnarray}
and it is not orthogonal to the transferred four momentum
$q=k-k^\prime$ even for massless neutrinos, i.e,
${ L}_{\mu\sigma}^{(\nu)} q^\mu = -m^2_l k_\sigma$.
%
%
%
\begin{figure}[tbh]
\centerline{\includegraphics[height=7cm]{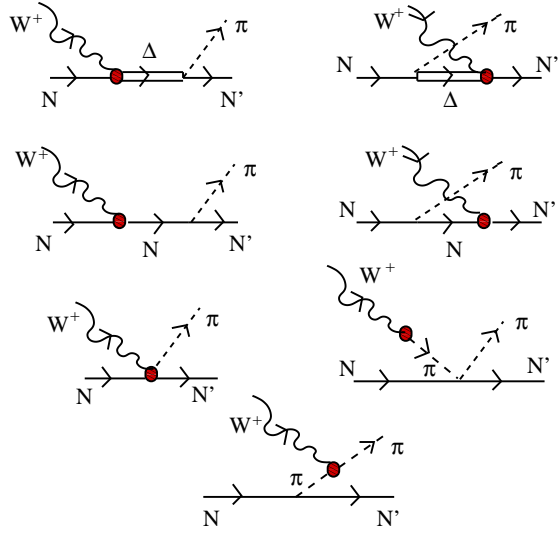}}
\caption{\footnotesize Model of Ref.~\cite{Hernandez:2007qq} for the $W^+N\to
  N^\prime\pi$ reaction.  The circle in the diagrams stands for the
  weak transition vertex. }\label{fig:diagramas}
\end{figure}

The hadronic tensor includes all the nuclear effects and it can be
approximated by
\begin{gather}
  { W}^{\mu\sigma}_{{\rm CC}\pi^+} =
  \frac{|\vec{k}_\pi|}{64\pi^3M^2}\, {\cal A}^\mu_{\pi^+}(q,k_\pi)
\left({\cal A}^\sigma_{\pi^+}(q,k_\pi)\right)^*
\label{eq:Jmunu}\\
{\cal A}^\mu_{\pi^+}(q,k_\pi) =  \int d^3\vec{r}\ e^{{\rm
        i}\left(\vec{q}-\vec{k}_\pi\right)\cdot\vec{r}}
    \left\{\rho_p(\vec{r}\,) 
\Big[{\cal J}^\mu_{p\pi^+}(\vec{r};q,k_\pi)\Big] 
+ \rho_n(\vec{r}\,) 
\Big[{\cal J}^\mu_{n\pi^+}(\vec{r};q,k_\pi)\Big] \right \}  
\label{eq:Jmunu2}
\end{gather}
with $M$ the nucleon mass, $\rho_{p(n)}$ the nuclear
proton (neutron) density, normalized to the number of protons
(neutrons).  Since we have neglected the recoil energy of the final
nucleus we have $q^0=k_\pi^0(\equiv E_\pi)$. Finally ${\cal
  J}^\mu_{N\pi^+}(\vec{r};q,k_\pi)$, stands for the nucleon helicity
averaged $W^+ N \to N \pi^+$ amplitude evaluated inside the nuclear
medium as explained below.  Our model for the coherent nuclear
process is built up from the coherent scattering of the $W^+$ boson
with each of the nucleons of the nucleus producing an outgoing
$\pi^+$. The nucleon state (wave function) remains unchanged in the
dispersion and thus after summing over all nucleons, we obtain the
nuclear densities which appeared in the hadronic tensor ${
  W}^{\mu\sigma}_{{\rm CC}\pi^+}$ of Eq.~(\ref{eq:Jmunu}).  In the
elementary $W^+ N \to N \pi^+$ process, energy conservation is
accomplished by imposing $q^0=E_\pi$, while the transferred momentum
$\vec{q}-\vec{k}_\pi$ has to be accommodated by the nucleon wave
functions. Thus, the coherent pion production process is sensitive to
the Fourier transform of the nuclear density for momentum
$\vec{q}-\vec{k}_\pi$ (see Eq.~(\ref{eq:Jmunu2})).  This nuclear form
factor gets its maximum value when $\vec{q}$ and $\vec{k}_\pi$ are
parallel, but for this particular kinematics the vector contribution
of the ${\cal J}^\mu_{N\pi}$ amplitudes, which is purely transverse
$\vec{k}_\pi \times \vec{q}$\,,
vanishes. This is the reason why for electron and photon induced
reactions, the coherent pion production cross section turned out to be
a quite  small fraction of the total inclusive nuclear absorption
one~\cite{Carrasco:1991we,Hirenzaki:1993jc}. For neutrino induced
reactions, the axial contribution of the amplitudes is not suppressed
for kinematics where $\vec{q}$ and $\vec{k}_\pi$ are almost
parallel. Thus, the reduction induced by the nuclear form factor
is much less important, and one might expect a larger relative
contribution of the coherent pion production channel, as it is the
case for some purely hadron reactions (f.i., coherent pion production in
the ($^3$He,t) in nuclei~\cite{FernandezdeCordoba:1992ky}). This dominance of
the axial contributions has
been extensively exploited, through the PCAC hypothesis, to relate the
neutrino coherent pion production cross section with the pion-nucleus
elastic differential one~\cite{Rein:1982pf,Rein:2006di,Paschos:2005km}.

For the elementary process we have used the model recently
derived in Ref.~\cite{Hernandez:2007qq}. In addition to the $\Delta$ pole
($\Delta P$) mechanism (weak excitation of the $\Delta(1232)$
resonance and its subsequent decay into $N\pi$), the model also includes
background terms required by chiral symmetry.  It consists of seven
diagrams (see Fig.~\ref{fig:diagramas}): Direct and crossed
$\Delta(1232)-$ (first row) and nucleon (second row) pole terms
($\Delta P$, $C\Delta P$, $NP$, $CNP$) contact ($CT$) and pion pole
($PP$) contribution (third row) and finally the pion-in-flight ($PF$)
term. It provides a fairly good description of all available data for
pion production off the nucleon at intermediate energies, driven by CC
and NC and induced by both neutrino and antineutrino~\cite{Hernandez:2007qq}. To compute ${\cal
  J}^\mu_{N\pi^+}(\vec{r};q,k_\pi)$, we will need to evaluate
\begin{equation}
\frac12 \sum_r \bar u_r(\vec{p}^{\,\prime}) \Gamma_{i; N\pi^+}^\mu
u_r(\vec{p}\,),
\quad i=\Delta P,\, C\Delta P,\, NP,\, CNP,\, CT,\, PP,\, PF \label{eq:avg}
\end{equation}
where the $u$'s are Dirac spinors for the nucleons, normalized such
that $\bar u u=2M$, and the four-vector matrices $\Gamma_{i;
  N\pi^+}^\mu$ can be read from the explicit expressions of the pion
production amplitudes $\langle N\pi^+ | j^\mu_{cc+}(0)| N\rangle =
\sum_i \bar u(\vec{p}^{\,\prime}) \Gamma_{i; N\pi^+}^\mu u(\vec{p}\,)
$ in Eq.~(51) of Ref.~\cite{Hernandez:2007qq}. Finally, $\vec{p}$ and
$\vec{p}^{\,\prime}=\vec{p}+\vec{q}-\vec{k}_\pi$ are the initial and
final three momenta of the nucleon. Those momenta are not well defined
and we approximate the four-momentum of the nucleon $N$ which collides
with the $W^+$ by
\begin{equation}
p^\mu = \bigg( \sqrt{M^2+ \frac14{\left(\vec{k}_\pi-\vec{q}\right)^2}} ,
  \frac{\vec{k}_\pi-\vec{q}}{2}\,\bigg)  \label{eq:pmu} \, ,
\end{equation}
Hence we assume that the initial nucleon momentum is
$(\vec{k}_\pi-\vec{q}\,)/2$ and the final one is
$-(\vec{k}_\pi-\vec{q}\,)/2$, with both nucleons being on-shell.  The
momentum transfer is equally shared between the initial and final
nucleon momenta. This prescription was firstly used in
Refs.~\cite{Carrasco:1991we,Hirenzaki:1993jc} for coherent $\pi^0$
photo- and electroproduction, respectively. The approximation is based
on the fact that, for Gaussian nuclear wave functions, it leads to an
exact treatment of the terms linear in momentum of the elementary
amplitude. In Ref.~\cite{Carrasco:1991we} it was shown that this
prescription provided similar results as the explicit sum for the
nucleon momenta performed in Ref.~\cite{Boffi:1991nh}. More recently
it has also been employed in
Refs.~\cite{Peters:1998mb,Drechsel:1999vh} for coherent $\pi^0$ photo-
and electroproduction and in a recent work on neutrino coherent pion
production~\cite{AlvarezRuso:2007tt,AlvarezRuso:2007it}.
Setting $\vec{p}=-\vec{p}^{\,\prime}=(\vec{k}_\pi-\vec{q}\,)/2$, with
energies $p^0=p^{\prime 0}$ given by Eq.~(\ref{eq:pmu}), eliminates
some non-local contributions, and it greatly simplifies the sum over all
nucleons, which can be cast in terms of the neutron and proton
densities (see Eq.~(\ref{eq:Jmunu2})). Furthermore, the sum over
helicities in Eq.~(\ref{eq:avg}) can be also easily performed for
$\vec{p}=-\vec{p}^{\,\prime}$ since
$u_r(\vec{p}^{\,\prime}=-\vec{p}\,)= \gamma^0 u_r(\vec{p}\,)$, so that
\begin{equation}
\frac12 \sum_r \bar u_r(\vec{p}^{\,\prime}=-\vec{p}\,) \Gamma_{i; N\pi^+}^\mu
u_r(\vec{p}) = \frac12 {\rm  Tr}\left((\slashchar{p}+M)\gamma^0\Gamma_{i; N\pi^+}^\mu \right),
\quad i=\Delta P\,, C\Delta P\,, NP\ldots \label{eq:avg2}
\end{equation}
Thus finally, we will use
\begin{gather} 
{\cal J}^\mu_{N\pi^+}(\vec{r};q,k_\pi) = 
\sum_i {\cal J}^\mu_{i;N\pi^+}(\vec{r};q,k_\pi), 
\quad i=\Delta P,\, C\Delta P,\, NP,\, CNP,\, CT,\, PP,\, PF \\
{\cal J}^\mu_{i;N\pi^+}(\vec{r};q,k_\pi) =
\frac12 {\rm Tr}\left((\slashchar{p}+M)\gamma^0 \Gamma_{i; N\pi^+}^\mu
\right)\frac{M}{p^0} \, ,
\end{gather}
expressions that we shall evaluate numerically. Note that the $PF$
term does not contribute to the process since the trace above is zero
in this case. Within this approximation, the averaged $W^+ N \to N
\pi^+$ amplitude inside the nuclear medium, ${\cal
  J}^\mu_{N\pi^+}(\vec{r};q,k_\pi)$, does not depend on $\vec{r}$.
Below, we will include further medium corrections to the dominant
$\Delta P$ mechanism which will induce an explicit $\vec{r}$
dependence.

Given the importance of the $\Delta-$pole contribution and since the
$\Delta$ properties are strongly modified inside the nuclear
medium~\cite{Hirata:1978wp,Oset:1981ih,Oset:1987re,Nieves:1993ev}, we
consider some additional nuclear corrections to this contribution to
include the effect of the self-energy of the $\Delta$ in the medium
$\Sigma_\Delta(\rho(\vec{r}\,))$. Here we follow the same approach as
in Ref.~\cite{AlvarezRuso:2007tt}, which is based on the findings of
Refs.~\cite{Oset:1987re,Nieves:1993ev,Nieves:1991ye} . Thus in the
$\Delta-$propagator, we make the substitutions $M_\Delta\to M_\Delta +
{\rm Re}\Sigma_\Delta $ and $\Gamma_\Delta/2 \to \Gamma_\Delta^{\rm
  Pauli}/2- {\rm Im} \Sigma_\Delta$ and take
$\Sigma_\Delta(\rho(\vec{r}\,))$ and $\Gamma_\Delta^{\rm Pauli}/2$ as
explained in Sect. II-B of Ref.~\cite{AlvarezRuso:2007tt}.

So far the formalism has used the bound wave functions of the nucleus,
which appear via the proton and neutron densities, and has considered
only a plane wave for the pion. Pion distortion effects are important,
specially for $|\vec{k}_\pi| < 0.5$ GeV~\cite{Kelkar:1996iv,Singh:2006bm,AlvarezRuso:2007tt,AlvarezRuso:2007it}, and are considered here by replacing in
Eq.~(\ref{eq:Jmunu2})
\begin{gather}
e^{-{\rm i}\vec{k}_\pi\cdot\vec{r}} \to  \widetilde{\varphi}_{\pi^+}^{\ast}
    (\vec{r};\vec{k}_\pi) \label{eq:pi-nolocal}  \\
\vec{k}_\pi e^{-{\rm i}\vec{k}_\pi\cdot\vec{r}} \to {\rm i}
\vec{\nabla} \widetilde{\varphi}_{\pi^+}^{\ast}
    (\vec{r};\vec{k}_\pi) \label{eq:pi-nolocal2}
\end{gather} 
The pion wave function $\widetilde{\varphi}_{\pi^+}^{\ast}
(\vec{r};\vec{k}_\pi)$  corresponds to an incoming solution
of the Klein Gordon equation,
%
\begin{equation}
\left [- \vec{\bigtriangledown}^{2} + m_\pi^{2} + 2 E_\pi V_{\rm opt}
(\vec{r})\right ] 
\widetilde{\varphi}_{\pi^+}^{\ast}
    (\vec{r};\vec{k}_\pi) =
E_\pi^{2} \widetilde{\varphi}_{\pi^+}^{\ast}
    (\vec{r};\vec{k}_\pi) \,, \label{eq:optical}
\end{equation}
with $V_{\rm opt} (\vec{r})$ the optical potential which describes the
$\pi^+$-nucleus interaction.  This potential has been developed
microscopically and it is explained in detail in
Refs.~\cite{Nieves:1993ev,Nieves:1991ye}. It contains the ordinary
lowest order optical potential pieces constructed from the $s$-- and
$p$--wave $\pi N$ amplitudes. In addition second order terms in both
$s$-- and $p$--waves, responsible for pion absorption, are also
considered.  Standard corrections, as second-order Pauli re-scattering
term, angular transform term (ATT), Lorentz--Lorenz effect and long
and short range nuclear correlations, are also taken into account.
This theoretical potential reproduces fairly well the data of pionic
atoms (binding energies and strong absorption
widths)~\cite{Nieves:1993ev} and low energy $\pi$--nucleus
scattering~\cite{Nieves:1991ye}. At low pion energies, 
it is an improvement over the one used
in \cite{AlvarezRuso:2007tt,AlvarezRuso:2007it}, that was based on
$\Delta$ dominance of the $\pi N$ interaction. Another possible
improvement would be the inclusion of the Coulomb interaction between
the outgoing pion and the nucleus. This can be taken into account by
means of the replacement
\begin{equation}
E_\pi \to E_\pi - V_C (\vec{r}) \label{eq:coulomb}
\end{equation}
in the right hand side of Eq.~(\ref{eq:optical}), where $V_C(\vec{r})$ is
the Coulomb potential created by the nucleus, including finite size
and vacuum polarization effects, see
\cite{Nieves:1991ye,Nieves:1993ev}. We will  discuss the effect
of this correction in Sect.~\ref{sec:results}

The replacement in Eq.~(\ref{eq:pi-nolocal2}), that takes into
account the fact that the pion three momentum is only well defined
asymptotically when the pion-nucleus potential vanishes, induces some
non-localities in the amplitudes. To treat these non-localities we
have adopted the following scheme:
\begin{itemize}
\item In Eq.~(\ref{eq:Jmunu}), we approximate $|\vec{k}_\pi|$, which
arises from the phase space integrations, by the modulus of the
asymptotic three momentum
$
(E_\pi^2-m_\pi^2)^\frac12$

\item In the $\Delta P$, $C\Delta P$, $NP$, $CNP$ terms, we note that
  there exist either a $NN\pi$ or a $N\Delta \pi$ vertex (see Eq.~(51)
  of Ref.~\cite{Hernandez:2007qq}), which induces a factor
  $k_\pi^\alpha$ in the amplitudes. Indeed, for those terms we could
  re-write
\begin{equation}
{\cal J}^\mu_{i;N\pi^+}(\vec{r};q,k_\pi) = 
(k_\pi)_\alpha {\cal \hat{J}}^{\mu\alpha}_{i;N\pi^+}(\vec{r};q,k_\pi) , 
\quad i=\Delta P, C\Delta P\,, NP\,, CNP\, . 
\label{eq:prescription}
\end{equation}
We do not consider any non-locality in the tensor ${\cal
\hat{J}}^{\mu\alpha}_{i;N\pi^+}$, and we use the prescription of
Eqs.~(\ref{eq:pi-nolocal}) and (\ref{eq:pi-nolocal2}) to account for
$\vec{k}_\pi$ in the contraction between $k_\pi^\alpha$ and ${\cal
\hat{J}}^{\mu\alpha}_{i;N\pi^+}$ in Eq.~(\ref{eq:prescription}). This
approach to treat the non-localities is equivalent to that assumed in
refs.~\cite{AlvarezRuso:2007tt,AlvarezRuso:2007it}.

\item We do not consider any non-locality for the $CT$ and $PP$ contributions.

\end{itemize}
Antineutrinos induce the coherent production of negatively  charged
pions. To study these processes, we use~\cite{Hernandez:2007qq} 
\begin{eqnarray}
{ L}_{\mu\sigma}^{(\bar\nu)} &=& { L}_{\sigma\mu}^{(\nu)} \label{eq:anti1}\\ 
{\cal J}^\mu_{p\pi^- [n\pi^-]}(\vec{r};q,k_\pi) &=& 
{\cal J}^\mu_{n\pi^+ [p\pi^+]}(\vec{r};q,k_\pi)  \label{eq:anti2}
\end{eqnarray}
and implement the appropriate changes in the pion-nucleus 
$V_{\rm opt}$ and $V_C$ potentials to properly account  for the 
distortion of the outgoing $\pi^-$~\cite{Nieves:1991ye}.

Differences between neutrino and antineutrino induced cross sections
are proportional to the interferences among the axial and vector  
current contributions. Since the latter ones are suppressed by the nuclear
form factor, as we discussed after Eq.~(\ref{eq:Jmunu2}), we expect
roughly similar neutrino and antineutrino cross sections. This will also be
 the case for the NC driven processes  studied in the
next section.

\section{NC neutrino and antineutrino induced reactions}
\label{sec:nc}

To extend the above formulae to the case of NC $\pi^0$ coherent
production,
\begin{equation}
  \nu_l (k) + A_Z|_{gs}(p_a)  \to  
\nu_l (k^\prime) + A_Z|_{gs}(p^\prime_a) + \pi^0(k_\pi) 
\label{eq:neureac}
\end{equation}
we have, 
\begin{gather}
\frac{d^{\,5}\sigma_{\nu \nu}}{d\Omega(\hat{k^\prime})dE^\prime
    d\Omega(\hat{k}_\pi) } =
    \frac{|\vec{k}^\prime|}{|\vec{k}~|}\frac{G^2}{16\pi^2}
    { L}_{\mu\sigma}^{(\nu)}\, 
{ W}^{\mu\sigma}_{{\rm NC}\pi^0}\label{eq:nc-sec} 
\\ { W}^{\mu\sigma}_{{\rm NC}\pi^0} =
    \frac{|\vec{k}_\pi|}{64\pi^3M^2} {\cal A}^\mu_{\pi^0}(q,k_\pi)
    \left({\cal A}^\sigma_{\pi^0}(q,k_\pi)\right)^* \label{eq:zmunu}
\\ 
{\cal A}^\mu_{\pi^0}(q,k_\pi) = 
\int d^3\vec{r}\ e^{{\rm i}\vec{q}\cdot\vec{r}} 
\left\{\rho_{p}(\vec{r}\,) \big[{\cal J}^\mu_{p\pi^0 }(\vec{r};q,k_\pi)\big] 
+ \rho_{n}(\vec{r}\,)
    \big[{\cal J}^\mu_{n\pi^0 }(\vec{r};q,k_\pi)\big] \right\}
    \widetilde{\varphi}_{\pi^0}^{\ast} (\vec{r};\vec{k}_\pi)
\end{gather}
with $\vec{k}^\prime~$ and $E^{\prime}= |\vec{k}^{\prime\,}|$ the LAB
outgoing neutrino momentum and energy. The leptonic tensor is given in
Eq.~(\ref{eq:lep}) and it is now orthogonal to
$q^\mu=(k-k^\prime)^\mu$ for massless neutrinos, i.e,
${ L}_{\mu\sigma}^{(\nu)} q^\mu = { L}_{\mu\sigma}^{(\nu)} q^\sigma = 0$.

Both lepton and hadron tensors are independent of the neutrino lepton
family, and therefore the cross section for the reaction of
Eq.~(\ref{eq:neureac}) is the same for electron, muon or tau incident
neutrinos. Furthermore, the hadron tensor is the same for neutrino and
antineutrino induced reactions, and thus to study antineutrino
reactions we just have to change the sign of the antisymmetric part of
the leptonic tensor (see Eq.~(\ref{eq:anti1})).

To evaluate the hadronic tensor, we use the model for the NC pion
production off the nucleon derived in Ref.~\cite{Hernandez:2007qq} and thus we
have
\begin{gather} 
{\cal J}^\mu_{N\pi^0}(\vec{r};q,k_\pi) = 
\sum_i {\cal J}^\mu_{i;N\pi^0}(\vec{r};q,k_\pi),\, 
\quad i=\Delta P,\, C\Delta P,\, NP,\, CNP,\, CT,\, PP,\, PF
\label{eq:cn-jcoh} \\
  {\cal J}^\mu_{i;N\pi^0}(\vec{r};q,k_\pi) = 
\frac12 {\rm Tr}\left((\slashchar{p}+M)\gamma^0\,\Gamma_{i;
  N\pi^0}^\mu \right)\frac{M}{p^0}  \, .
\label{eq:cn-jcoh2}
\end{gather}
with the same prescription for the nucleon momentum as in the CC case. 
Within this model the PP, PF and CT diagrams do not contribute to the
NC $\pi^0$ production off the nucleon. The $\Delta P$ and $C\Delta P$
terms provide equal $Z^0 p \to p \pi^0$ and $Z^0 n \to n \pi^0$
amplitudes, with $\Gamma_{\Delta P; N\pi^0}^\mu$ and 
$\Gamma_{C\Delta P; N\pi^0}^\mu$ obtained from 
$\Gamma_{\Delta P;p\pi^+}^\mu$ and
$\Gamma_{C\Delta P;p\pi^+}^\mu$ multiplying these latter matrices by
the overall factors $2\sqrt 2/(3\cos\theta_C)$ and 
$2\sqrt 2/\cos\theta_C$ respectively, and multiplying the vector form factors
by $(1-2\sin^2\theta_W)$, being $\theta_C$ the Cabibbo angle and
$\theta_W$ the Weinberg angle. Direct and crossed nucleon pole terms
lead to
\begin{eqnarray}
\Gamma_{NP; N\pi^0}^\mu &=& 
-{\rm i}\,D^{NP}\frac{g_A}{2 f_\pi}
 \slashchar{k}_\pi\gamma_5
\frac{\slashchar{p}+\slashchar{q}+M}{(p+q)^2-M^2+ i\epsilon}
\left(V^{\mu}_{Z;N}(q) -A^\mu_{Z;N }(q) \right)  \,,
\quad D^{NP} = \left(\begin{array}{cc} \phantom{-}1
  & p\pi^0 \cr -1
  & n\pi^0 \end{array} \right ) 
\end{eqnarray}
\begin{eqnarray}
\Gamma_{CNP; N\pi^0}^\mu &=& 
-{\rm i}\,D^{CNP}\frac{g_A}{2 f_\pi}\left(V^\mu_{Z;N}(q)-A^\mu_{Z;N}(q)\right)
\frac{\slashchar{p}-\slashchar{k}_\pi+M}{(p-k_\pi)^2-M^2+ i\epsilon} 
\slashchar{k}_\pi\gamma_5,\quad D^{CNP} = \left(\begin{array}{cc} \phantom{-}1
  & p\pi^0 \cr -1
  & n\pi^0 \end{array} \right )
\end{eqnarray}  
with $f_\pi \simeq 93$ MeV the pion weak decay constant, $g_A=1.26$ the
axial nucleon coupling, and 
\begin{eqnarray}
V^\alpha_{Z;N} &=& 2\times \left [ F_1^{Z}(q^2)\gamma^\alpha + i 
 \frac{\mu_{Z} F_2^{Z}(q^2)}{2M}\sigma^{\alpha\nu}q_\nu\right]_N, \\
A^\alpha_{Z;N} &=& 
 \left[G_A^{Z}(q^2) \gamma^\alpha\gamma_5  + 
\left((G_A^{Z}(q^2) + G_A^s(q^2))
\frac{\slashchar{q}}{m_\pi^2-q^2}+G_P^s(q^2)\right)
q^\alpha\gamma_5\right]_N
\end{eqnarray}
The pseudoscalar part of the axial current, which is proportional to
$q^\mu$, does not contribute to the differential cross section for
massless neutrinos. Besides the $Z^0NN$ form factors are given
by~\cite{Nieves:2005rq}
\begin{eqnarray}
\left(F_1^Z\right)^{p,n} &=& \pm F_1^V - 2\sin^2\theta_W F_1^{p,n}-
\frac12 F_1^s \\
\left(\mu_{Z} F_2^Z\right)^{p,n} &=& \pm \mu_{V} F_2^V - 2\sin^2\theta_W 
\mu_{p,n}F_2^{p,n}- \frac12 \mu_s F_2^s \\
\left(G_A^Z\right)^{p,n} &=& \pm G_A - G_A^s   \, .
\end{eqnarray}
For isoscalar nuclei, the direct and crossed nucleon pole terms do not
contribute because the existing cancellation between neutron and
proton contributions, and we have total dominance of the $\Delta$
mechanisms. If we neglect the vector current
contributions\footnote{They will be suppressed by the nuclear form
factor (see discussion after Eq.~(\ref{eq:Jmunu2})).}, finite lepton
mass effects and approximating $\cos\theta_C\approx 1$, we find that
the CC coherent pion production cross section is twice the NC one, as
deduced from the relevant isospin factors and the factor of four of
difference between Eqs.~(\ref{eq:sec}) and (\ref{eq:nc-sec}), for CC
and NC driven processes, respectively\footnote{For the case of
isoscalar nuclei, the approximate relation $\sigma_{CC}\approx
2 \sigma_{NC}$ is far more general and  it can be directly deduced from
PCAC and isospin invariance.}. For non symmetric nuclei, as long as
the $\Delta$ dominance holds, we will reach the same
conclusion. Nevertheless, we remind here that for low and intermediate
muon neutrino energies ($\le 1.5-2 $ GeV), one should expect sizable
corrections (25\% at 1.3 GeV, and greater at smaller energies) to the
approximate relation $\sigma_{\rm CC} \approx 2 \sigma_{\rm NC}$ due
to the finite muon mass~\cite{Rein:2006di}.

To evaluate pion distortion effects we compute the $\pi^0-$ wave
function by using the appropriate pion-nucleus optical
potential~\cite{Nieves:1991ye} and setting $V_C=0$.  Non-localities in
the amplitudes are treated as in the CC case.

\section{Results}\label{sec:results}

We shall always use the full model of Ref.~\cite{Hernandez:2007qq}
where the dominant $C_5^A$ nucleon-to-$\Delta$ axial form factor was
fitted to data resulting in $C_5^A(0)=0.867$ and $M_{A\Delta}= 0.985$
GeV. Note that the Goldberger--Treiman relation, traditionally assumed
in the literature, implies a  larger value of $C_5^A(0)\sim
1.2$. We will come back to this point below, at the end of
the discussion of the results shown in Fig.~\ref{fig:background}. 

Firstly, we compile in Table~\ref{tab:dens} the input charge
densities, taken from Ref.~\cite{De Jager:1974dg}, used in this
work. For each nucleus we take the neutron matter density
approximately equal (but normalized to the number of neutrons) to the
charge density, though we consider small changes, inspired by
Hartree-Fock calculations with the density-matrix
expansion~\cite{Negele:1975zz} and corroborated by pionic atom
data~\cite{GarciaRecio:1991wk}. However, charge (neutron) matter
densities do not correspond to proton (neutron) point-like densities
because of the finite size of the nucleon. This is taken into account
by following the procedure outlined in section 2 of
Ref.~\cite{GarciaRecio:1991wk} (see Eqs. (12)-(14) of this reference).
\begin{table}
 \begin{center} \begin{tabular}{llll}\hline\tstrut 
 Nucleus & $r_{p}$ [fm]& $r_{n} $[fm] & \hspace*{.25cm}$a$ 
    \\\hline \tstrut
 $^{12}$C &1.692 & 1.692 & 1.082 \\
 $^{16}$O &1.833 & 1.833 & 1.544 \\
 $^{208}$Pb &6.624 & 6.890 & 0.549 fm \\
\hline \end{tabular}
 \end{center} 
\caption{ \footnotesize Charge ($r_p, a$) and neutron matter
 ($r_n, a$) density parameters for different nuclei as given in Ref.~\cite{De Jager:1974dg}. 
For carbon and oxygen we use a modified harmonic
 oscillator density, $\rho(r) = \rho_0 (1+a(r/r_N)^2)\exp(-(r/r_N)^2)$,
 while for lead, we use a two-parameter Fermi
 distribution, $\rho(r) = \rho_0 /(1+\exp((r-r_N)/a))$. }  
\label{tab:dens} 
\end{table} 

\subsection{General Results}

\begin{figure}[htb]
\begin{center}
\makebox[0pt]{\includegraphics[scale=0.7]{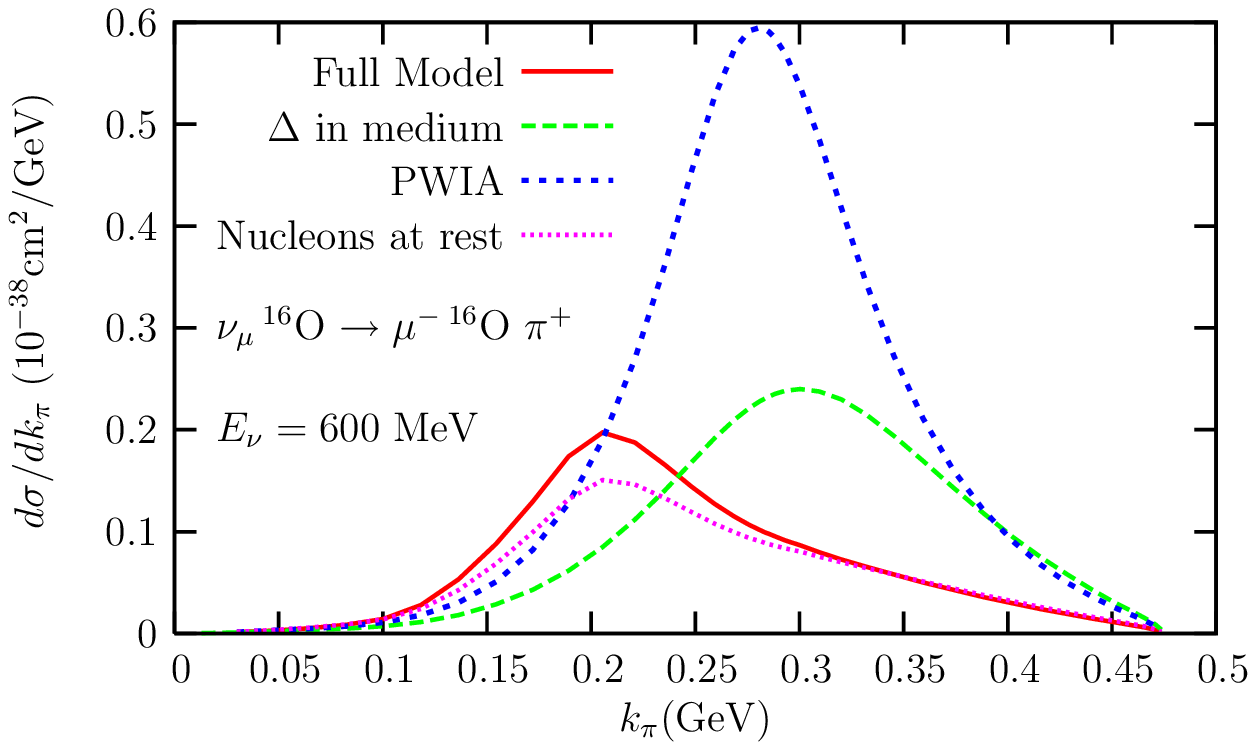}\hspace{0.5cm}
              \includegraphics[scale=0.7]{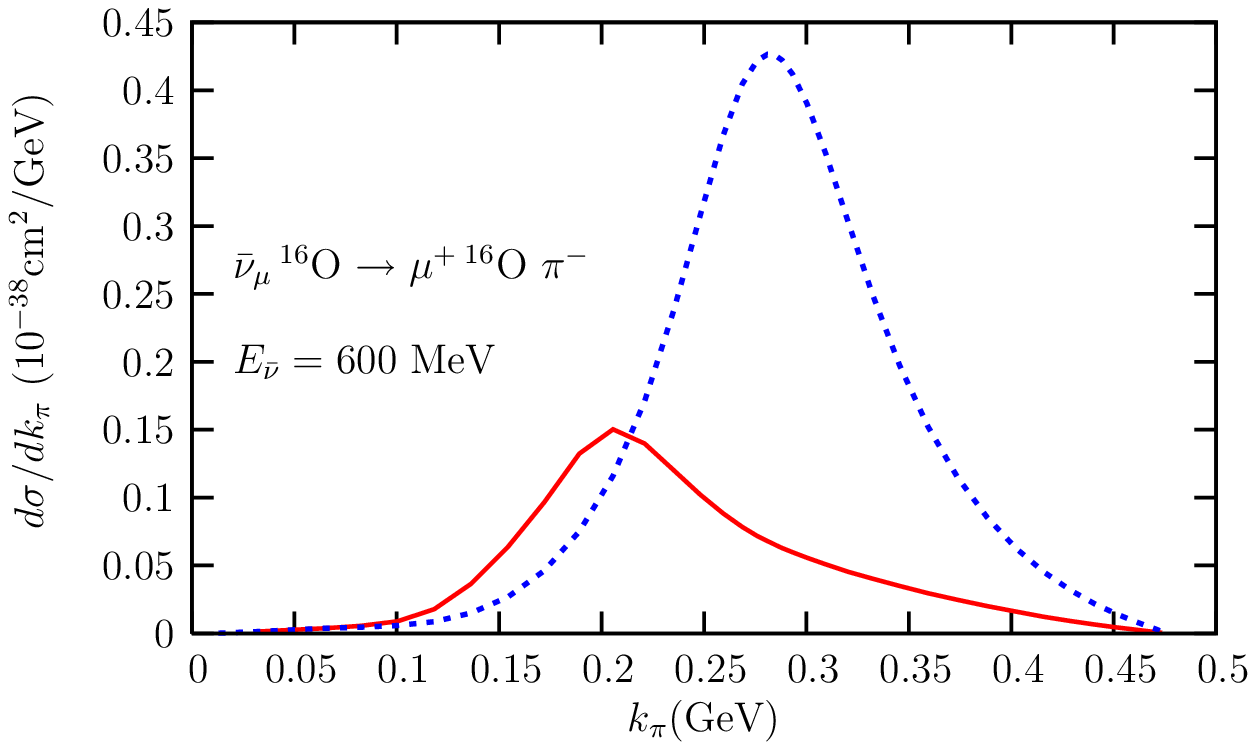}}
\\
\makebox[0pt]{\includegraphics[scale=0.7]{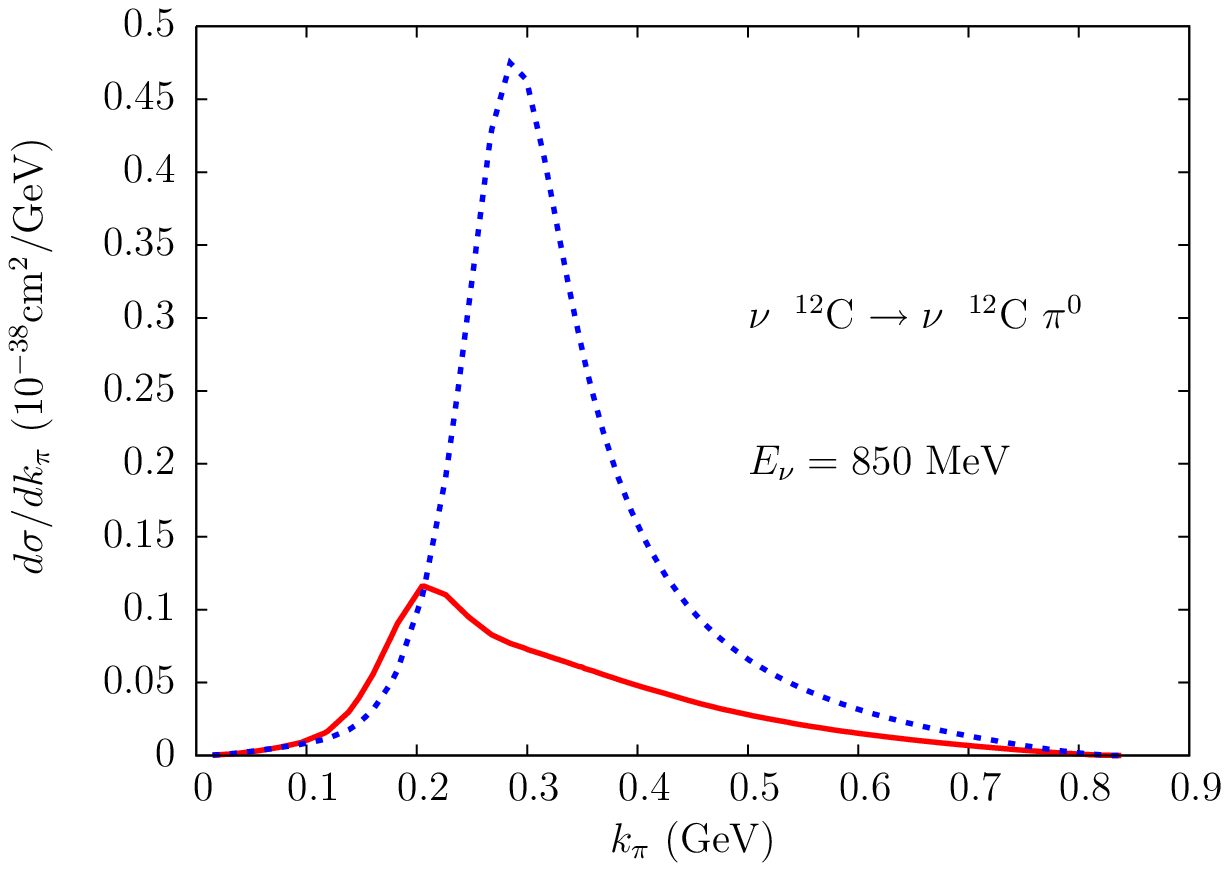}\hspace{0.5cm}
              \includegraphics[scale=0.7]{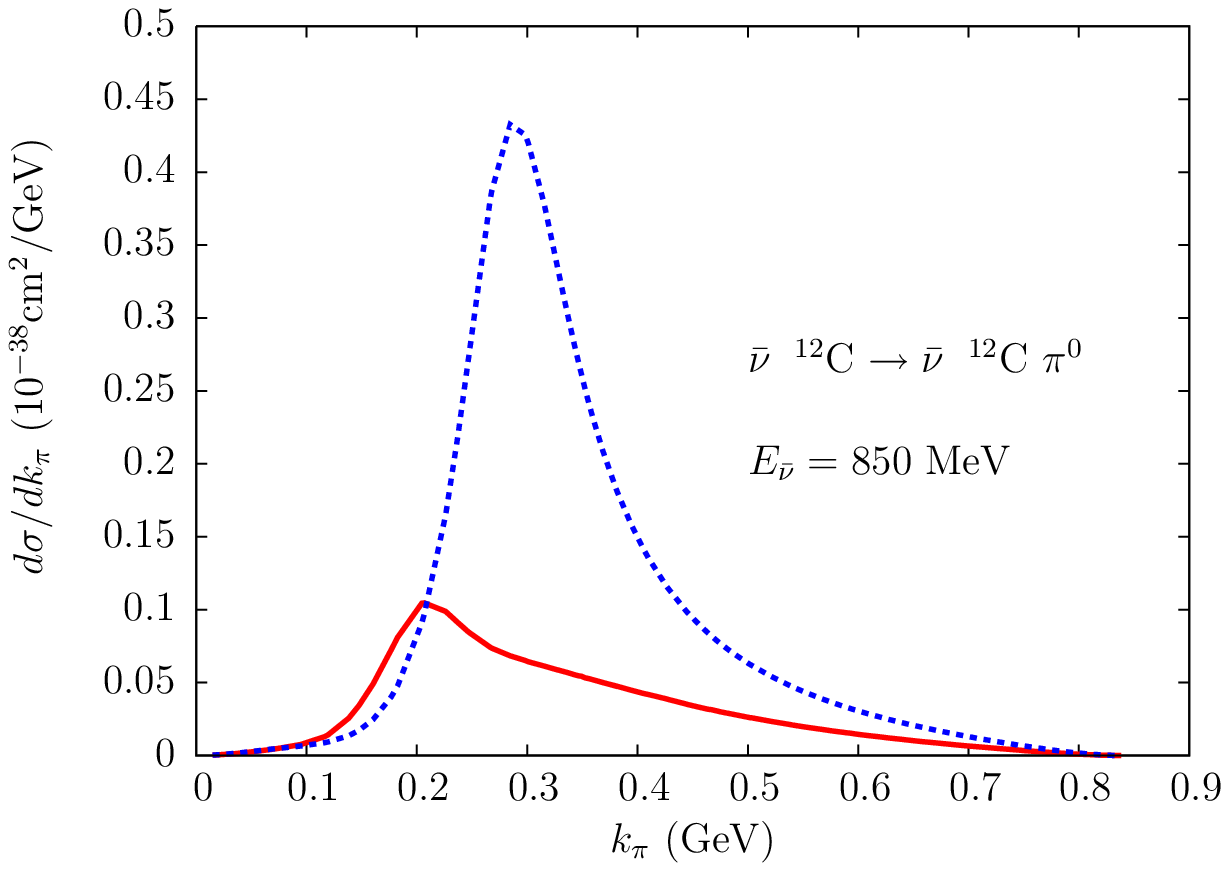}}
\end{center}
\caption{\footnotesize Pion momentum differential
 cross section in the LAB frame for
  different coherent pion production reactions.
  Short-dashed line (in blue) has been calculated using planes waves
  for the outgoing pion and without including any in-medium correction
  for the $\Delta$.  Results with $\Delta$ nuclear medium effects are
  shown in the upper-left panel by the long-dashed line (in green).
  Our full model calculation, including medium effects on the $\Delta$
  and the distortion of the outgoing pion wave function, is shown by
  the solid line (in red).  Finally, the effect of putting the
  nucleons at rest is shown in the upper-left panel by the dotted line
  (in magenta).  }\label{fig:dsdppi}
\end{figure}

First, in Fig.~\ref{fig:dsdppi}  we show the pion momentum distribution (LAB)
for CC and NC coherent pion production induced by $\nu_\mu$ and
$\bar\nu_\mu$ on a $^{16}$O (CC case) and $^{12}$C (NC case)
targets. In the upper panels we show the CC case for a
$\nu_\mu,\,\bar\nu_\mu$ beam energy of $600$ MeV, which is in the
expected peak energy region of the future T2K experiment. In the lower
two panels we show NC results for a $\nu\,\bar\nu$ beam energy of
$850$ MeV. In all panels, the short-dashed line corresponds to our
results in plane wave impulse approximation (PWIA), in which we use a
plane wave for the outgoing pion and neglect medium effects on the
dominant $\Delta$ contribution.  Medium effects introduced through the
$\Delta$ self-energy play a very important role largely reducing the
PWIA results. This is shown by the long-dashed line in the upper-left
panel.  When the pion distortion is also taken into account (via the
substitutions of Eqs.~(\ref{eq:pi-nolocal}) and
(\ref{eq:pi-nolocal2})) the cross section is further reduced, and the
peak is shifted towards lower energies reflecting the strong
absorption and the higher probability of a quasielastic collision of
the outgoing pion by the nucleus in the $\Delta$ kinematical
region. The total cross section reduction is around $60\% $ for a beam
energy of 600\,MeV. Our full model results thus obtained are shown by
the solid line. Medium and pion distortion effects in coherent pion
production were already evaluated in
Refs.~\cite{AlvarezRuso:2007tt,AlvarezRuso:2007it}. However in these
works the  motion of the nucleon was neglected. This is to say,
though the authors of these references also use the prescription
$\vec{p}=-\vec{p}^{\,\prime}=(\vec{k}_\pi-\vec{q}\,)/2$ to compute the
elementary $W^\pm(Z^0) N \to \pi N' $ amplitude, however the nucleon
momenta in the Dirac's spinors appearing in Eq.~(\ref{eq:avg2}) are
neglected. The effect of putting the nucleons at rest can be clearly
seen in the dotted line of the upper-left panel and results in a $\sim
15$\% decrease of the total cross section. Results obtained for the
antineutrino CC induced reaction are shown in the upper-right panel.
The cross section is some 30\% smaller than in the neutrino case due
to the sign change in the axial part of the lepton tensor which
results in a different vector-axial interference.  Similar effects are
seen for NC reactions shown in the lower panels of
Fig.~\ref{fig:dsdppi}. In this case the antineutrino cross section is
reduced by just 10\% with respect to the neutrino one.

\begin{figure}[htb]
\begin{center}
\makebox[0pt]{\includegraphics[scale=0.7]{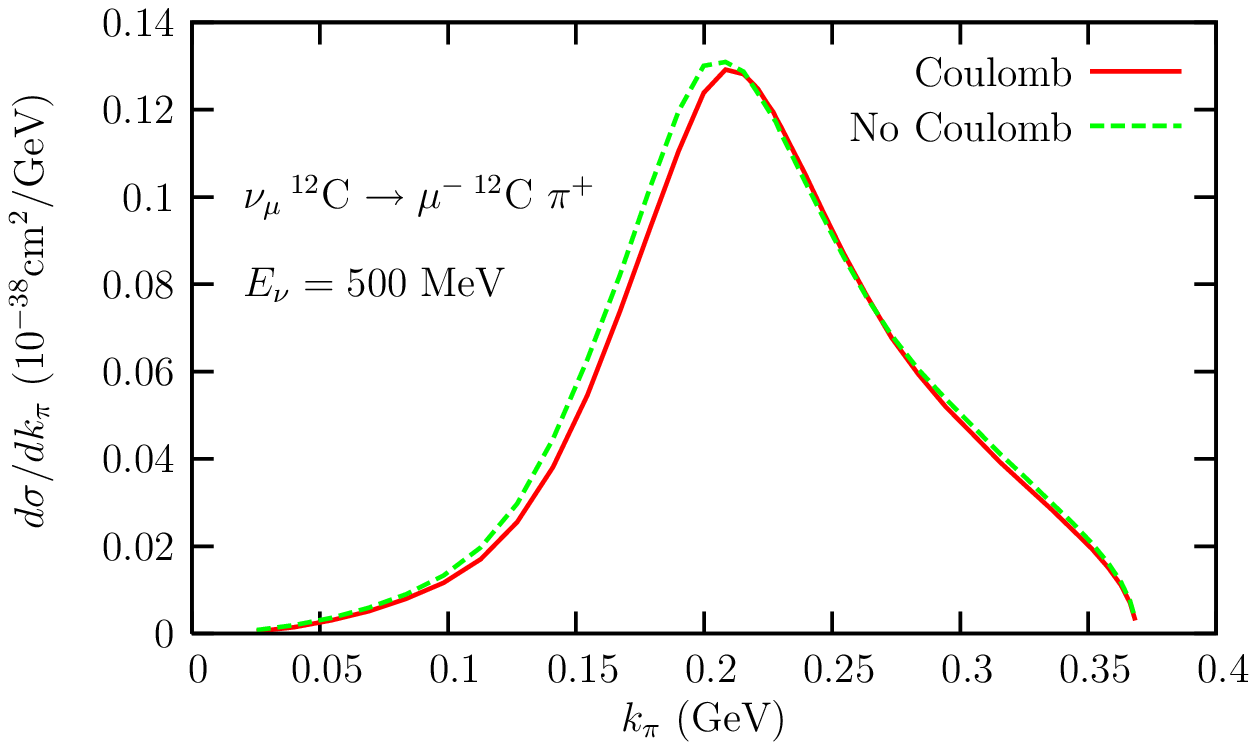}\hspace{0.5cm}
              \includegraphics[scale=0.7]{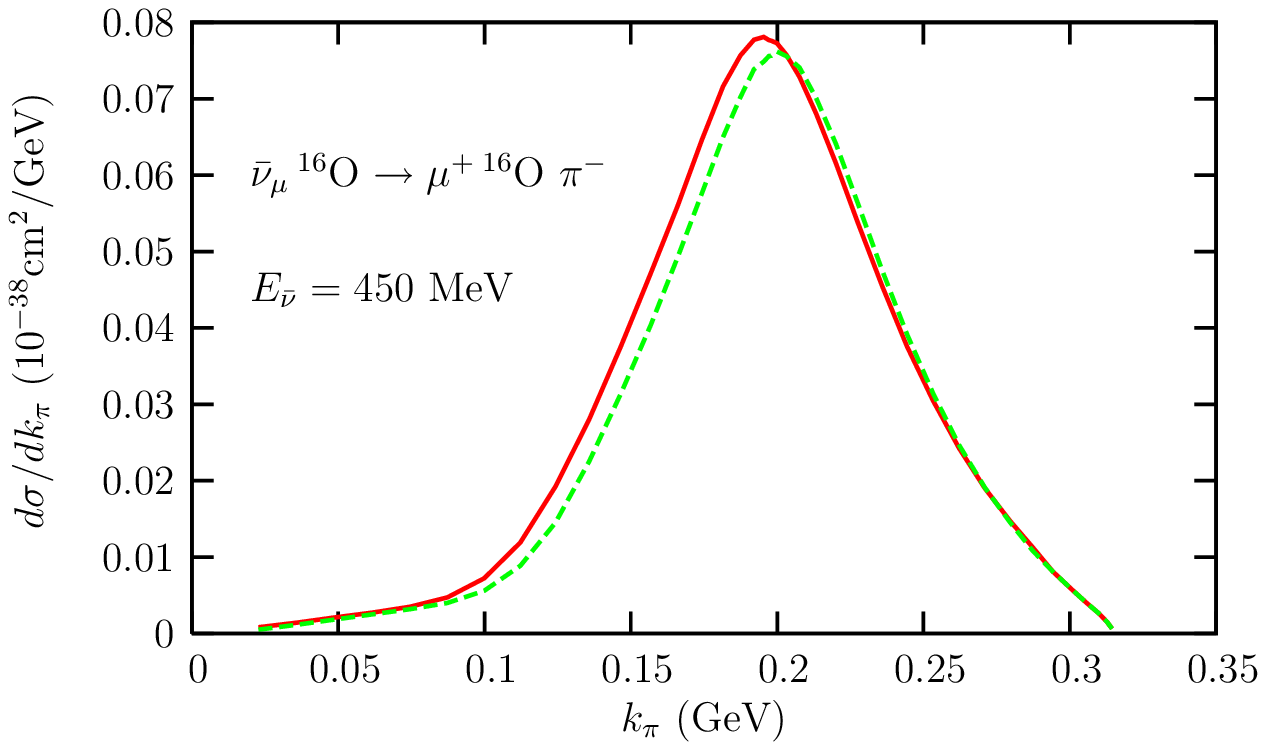}}
\end{center}
\caption{\footnotesize Coulomb distortion effects on the pion momentum
  differential LAB cross section for CC coherent pion
reactions induced by $\nu_\mu,\,\bar\nu_\mu$.} 
\label{fig:coulomb}
\end{figure}

In Fig.~\ref{fig:coulomb} we show the effect of Coulomb distortion on
the outgoing charged pion.  One can see the expected shift in the peak
towards higher (lower) energies of the positive (negative) pion
distribution when the Coulomb distortion is taken into account. The
net effect in the total cross section is nevertheless small, amounting
to a $5\% 
$ change for beam energies in the 500\,MeV region. For
higher energies the effect is expected to be less important.

\begin{figure}[htb]
\begin{center}
\makebox[0pt]{\includegraphics[scale=0.7]{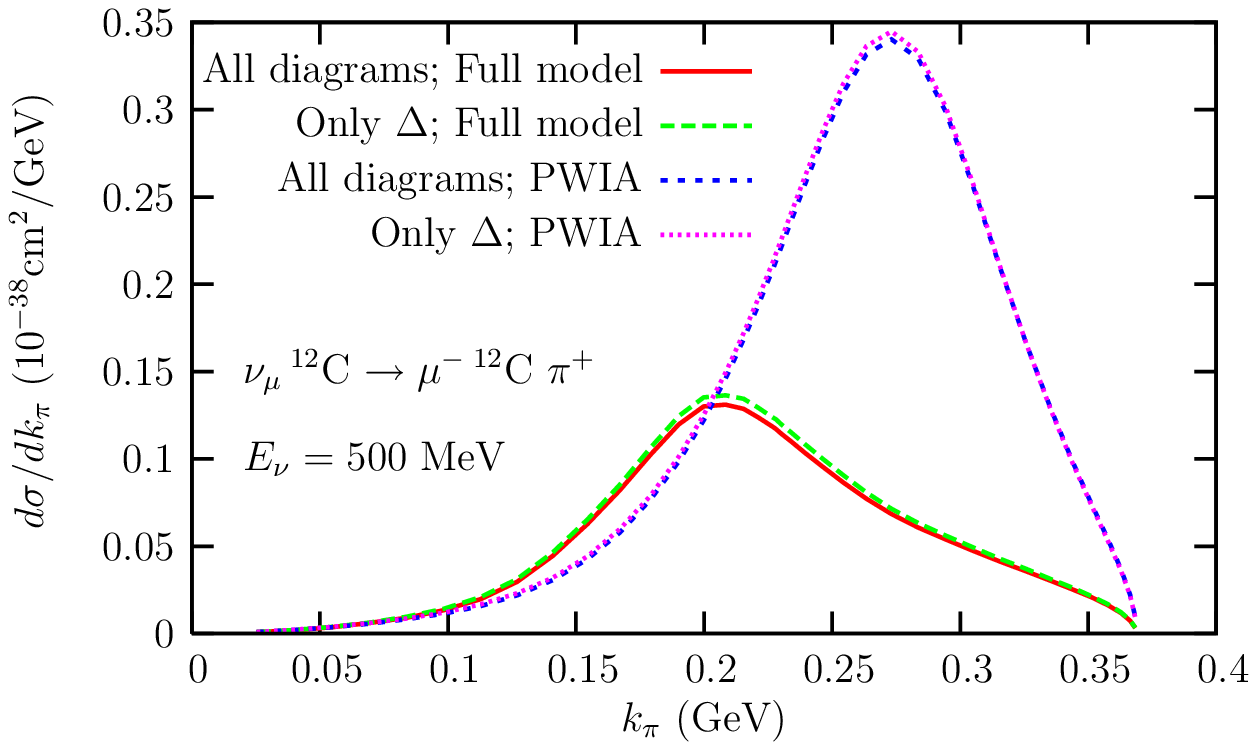}\hspace{0.5cm}
              \includegraphics[scale=0.7]{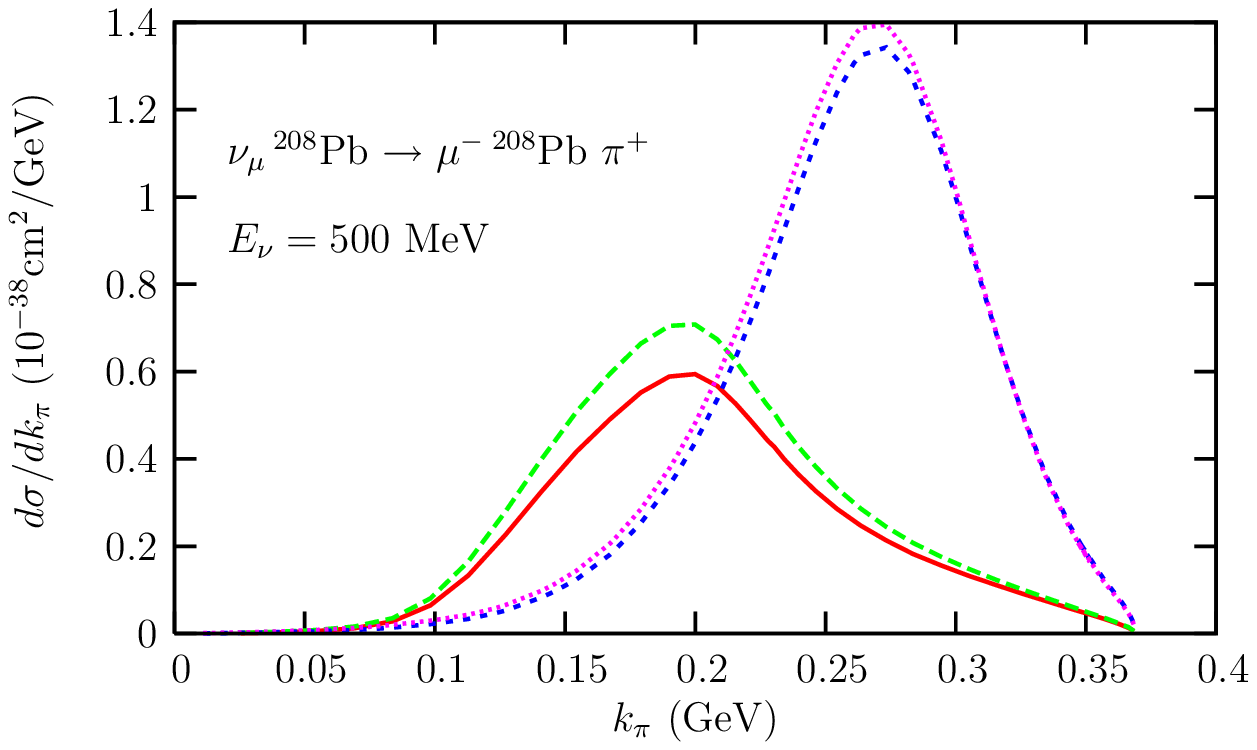}}
\end{center}
\caption{\footnotesize Pion momentum differential LAB cross section,
  calculated with and without background terms, for CC coherent pion
  production induced by a $500$ MeV $\nu_\mu$  beam. The curves
  labeled as ``Only $\Delta$'' stand for results obtained from the
  $\Delta P$ and $C\Delta P$ mechanisms. Left
  panel: results for a $^{12}$C target and PWIA and full
  calculations; Right panel: results for a $^{208}$Pb target.}
\label{fig:background}
\end{figure}

We now study, for CC reactions, the effect of including the background
terms on top of the dominant direct ($\Delta P$) and crossed ($C\Delta
P$) $\Delta$ pole contributions.  As mentioned before, the $PF$ term
does not contribute to the coherent cross section. Besides, the direct
($NP$) and crossed ($CNP$) nucleon pole term contributions partially
cancel each other, while the chiral background terms $CT$, $PP$ vanish
for isospin symmetric nuclei due to an exact cancellation between proton and
neutron contributions.  This latter cancellation is partial for
asymmetric nuclei. In the left panel of Fig.~\ref{fig:background} we
show the results for $^{12}$C. As seen
in the figure the effect of the background terms, both in the PWIA and
in the full calculation (including medium effects and pion
distortion), are very small, thus corroborating the findings of
Ref.~\cite{AlvarezRuso:2007it}.  In the right panel of the figure we
show full calculation results for $^{208}$Pb , which is the most
asymmetric nucleus with possible experimental interest. In this latter
case the inclusion of the background terms reduces the cross section
in an appreciable way. We find similar conclusions for NC driven
processes.

This predominant role played by the $\Delta$ mechanism, in conjunction
with the findings of Ref.~\cite{Hernandez:2007qq} and the fact that
the coherent pion production reaction in nuclei is mostly driven by
the axial part of the interaction, allows us to conclude that most of
the previous theoretical studies~\cite{Rein:1982pf, Rein:2006di,
Paschos:2005km, Singh:2006bm, AlvarezRuso:2007tt} of the pion coherent
channel might be overestimating the cross section roughly by a factor
of two. This can be easily understood as follows.  Background terms
turned out to be very important at the nucleon level and because of
them, the flux-averaged $\nu_\mu p\to \mu^- p \pi^+$ ANL
$q^2$-differential cross section ~\cite{Barish:1978pj,Radecky:1981fn}
is described with an axial form factor $C_5^A(0)$ of around
$0.9$~\cite{Hernandez:2007qq}, significantly smaller than the
traditionally used value of about $1.2$, deduced from the
Goldberger--Treiman relation. This reduction of the contribution of
the $\Delta$ pole mechanism in the weak pion production off the
nucleon is compensated by the non-resonant terms.  However, when one
studies the neutrino coherent pion production in isoscalar nuclei we
find a negligible contribution of the non-resonant terms, and thus the
cross section is determined by the axial part of the $\Delta$
mechanism of which $C_5^A$ gives the largest contribution. Thus, we
predict cross sections around a factor of $(1.2/0.9)^2\sim 2$ smaller
than those approaches which assume the Goldberger--Treiman
relation\footnote{A word of caution is required here.  There exists
some degree of inconsistency among the
ANL~\cite{Barish:1978pj,Radecky:1981fn} and BNL~\cite{bnl}
measurements of the integrated $\nu_\mu p\to \mu^- p \pi^+$ cross
section, being the latter larger than the former one.  The model of
Ref.~\cite{Hernandez:2007qq}, including non-resonant background terms,
with $C_5^A(0)=1.2$ (consistent with the off diagonal
Goldberger-Treiman relation) would lead to a better description of the
BNL data than if the lower vale of around 0.9 is used. Thus, if one
favours BNL data one could still use a high value of $C_5^A(0)=1.2$, 
which would lead to larger coherent pion production cross sections.}.
This fact was firstly pointed out by the authors of
Ref.~\cite{AlvarezRuso:2007it}, who used for the very first time the
background terms derived in Ref.~\cite{Hernandez:2007qq} for the
elementary reaction. However, we improve here the results of this
reference by properly  taking into account the  motion of the
nucleons, as discussed above, and correcting for some numerical
inaccuracies~\cite{luis} (of the order of 20\% in the total cross
section, and larger at the peak of the $d\sigma/dk_\pi$ differential
distribution) that affected the calculations of this reference and
those of a previous work by the same
authors~\cite{AlvarezRuso:2007tt}.

\begin{figure}[htb]
\begin{center}
\includegraphics[scale=0.7]{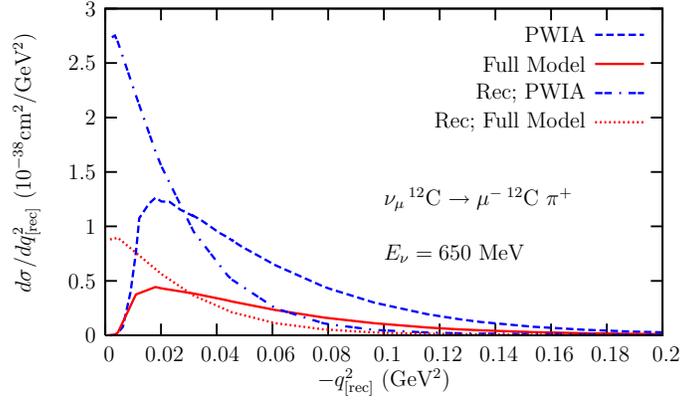}
\end{center}
\caption{\footnotesize Differential $d\sigma/dq^2$ and
LAB $d\sigma/dq^2_{\rm rec}$ distributions from carbon target, calculated
both within the PWIA and using our full model.  In the $x-$ axis,
$-q^2$ or LAB $-q^2_{\rm rec}$ values are displayed, depending on the curve. }\label{fig:fdq2_rec}
\end{figure}

Next we pay attention to the $q^2$ differential cross section, and in
Fig.~\ref{fig:fdq2_rec} we show this distribution for $\nu_\mu$ CC
driven processes in the energy region of the future T2K
experiment. There, we also show $d\sigma/dq^2_{\rm rec}$, where
$q^2_{\rm rec}$ is calculated, under the assumption of a Quasi-Elastic
(QE) neutrino--nucleon interaction, from the measured outgoing lepton
energy and scattering angle in the LAB frame,
\begin{eqnarray}
q^2_{\rm rec} = -2 M q^0_{\rm rec} &=& -2 M \left (E^{\rm rec}_\nu
-E^\prime\right ) \\
E^{\rm rec}_\nu &=& \frac{ME^\prime-m_l^2/2}{M-E^\prime+
  |\vec{k}^\prime| \cos\theta^\prime}
\end{eqnarray}
The $q^2_{\rm rec}$ distribution as compared with the $q^2$ one 
is clearly shifted to lower absolute values and peaks roughly at
0. This fact might be used to reduce the QE background by requiring
that coherent events  should have a reconstructed $q^2_{\rm rec}$ value smaller than some
appropriate cut, as was done in the K2K analysis carried out in
Ref.~\cite{Hasegawa:2005td}.

\begin{figure}[htb]
\begin{center}
\makebox[0pt]{\includegraphics[scale=0.7]{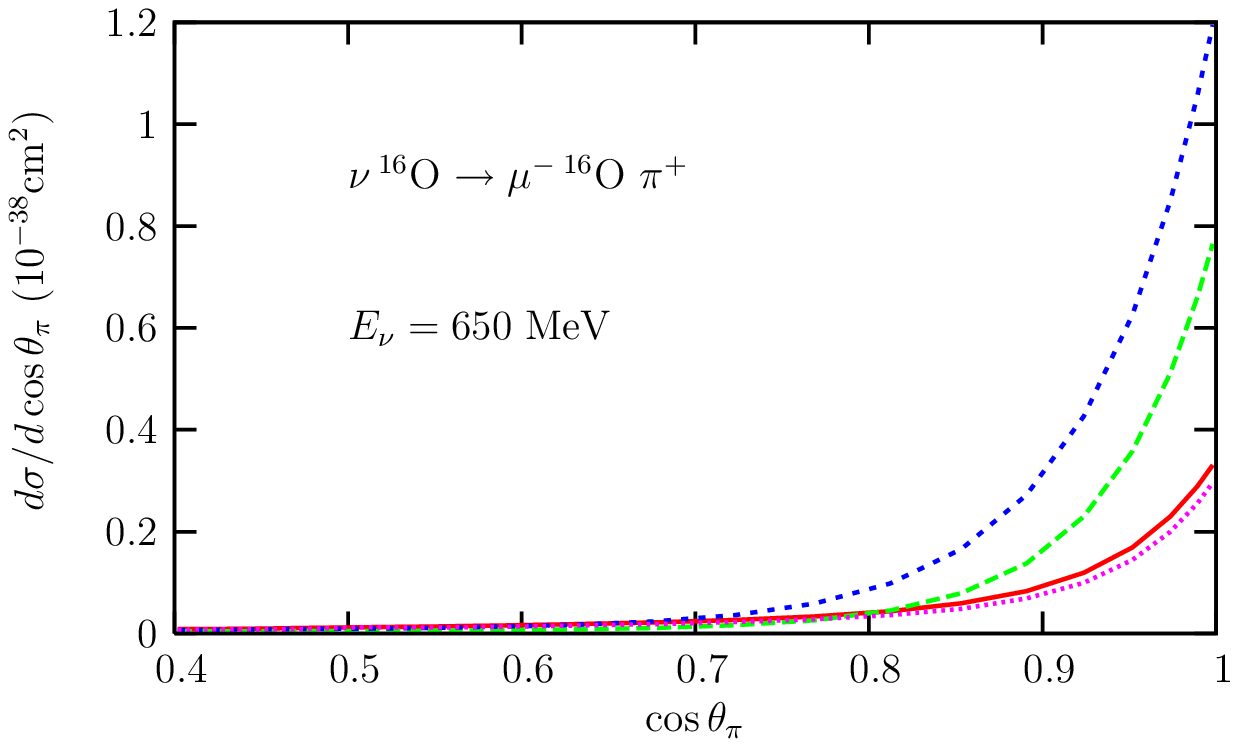}\hspace{0.5cm}
              \includegraphics[scale=0.7]{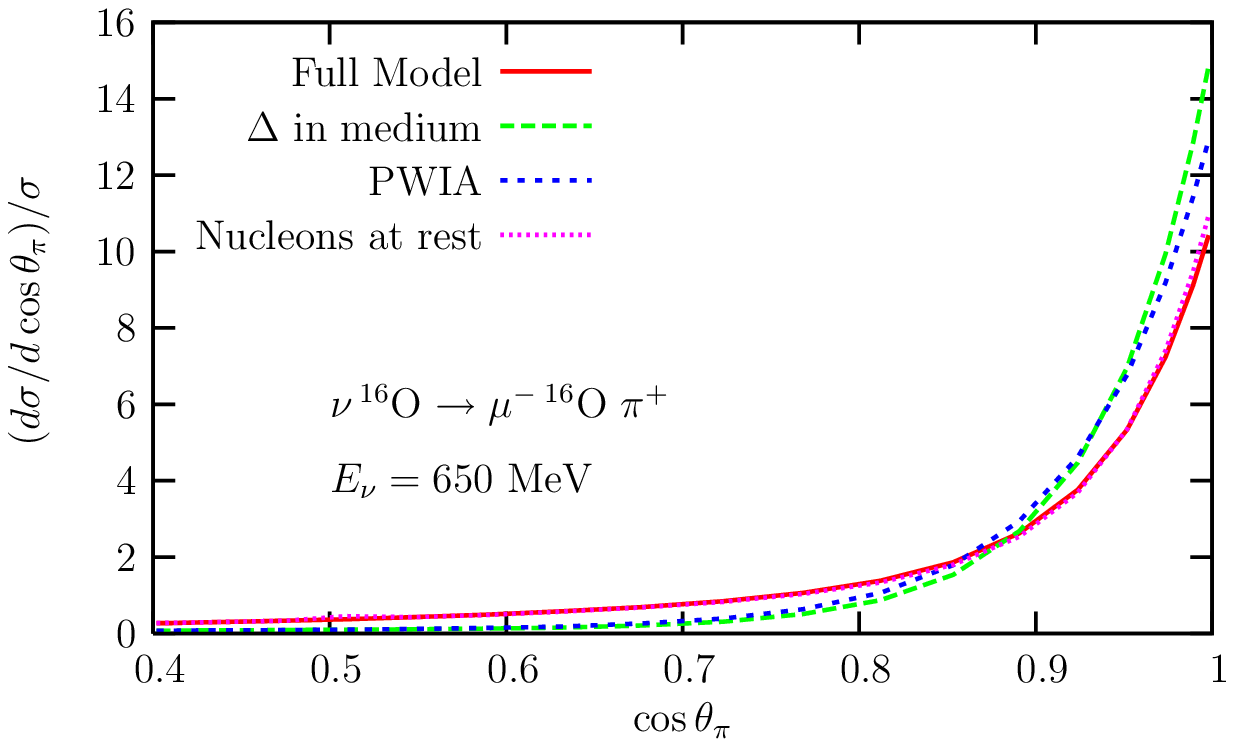}}
\end{center}
\caption{\footnotesize Pion angular differential  cross section for CC
  coherent pion production by a $\nu_\mu$ beam of $650$ MeV energy on
  a $^{16}$O target. Left panel: absolute differential cross section; Right
  panel: differential cross sections normalized to one. The
  $\theta_\pi$ angle is referred to the incoming neutrino
  direction in the LAB frame.}\label{fig:dsdcospi}
\end{figure}

\begin{figure}[htb]
\begin{center}
\makebox[0pt]{\includegraphics[scale=0.7]{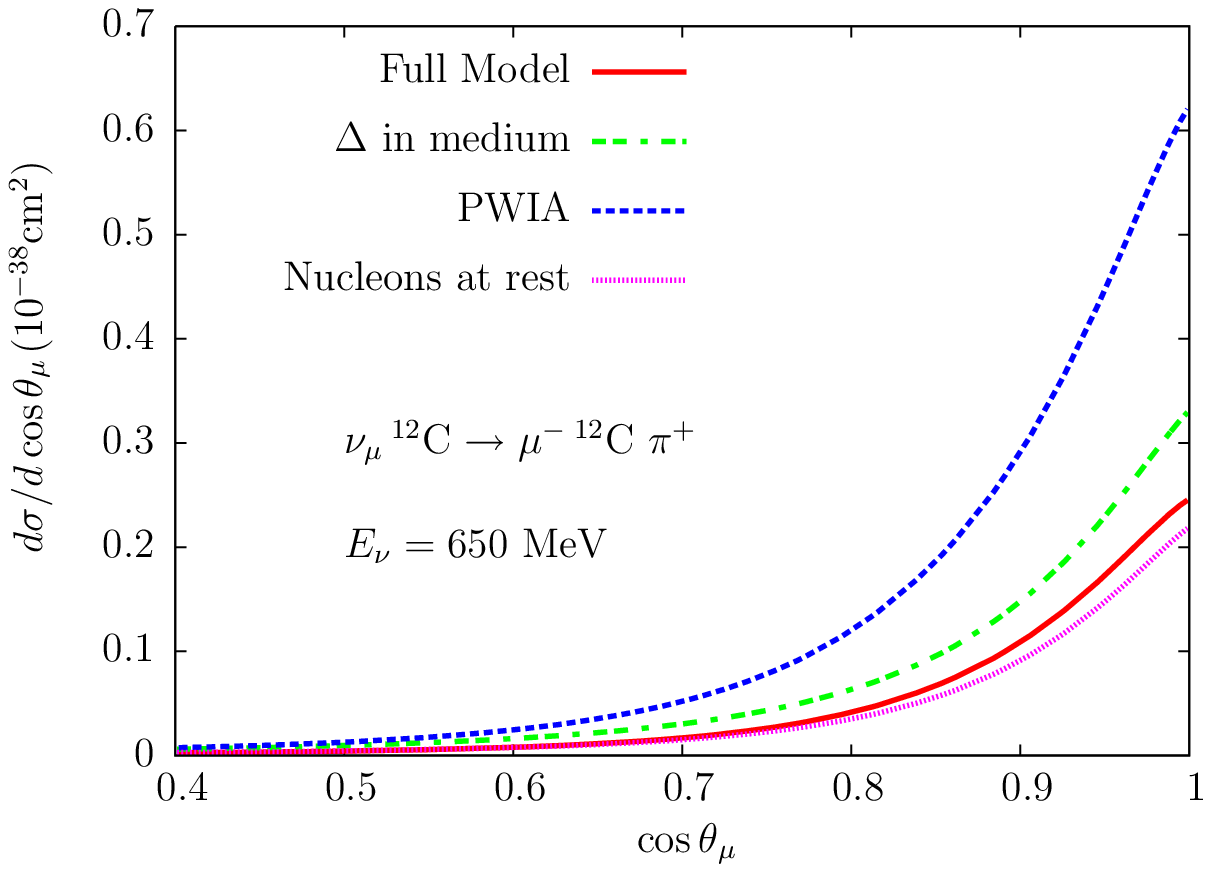}\hspace{0.5cm}
              \includegraphics[scale=0.7]{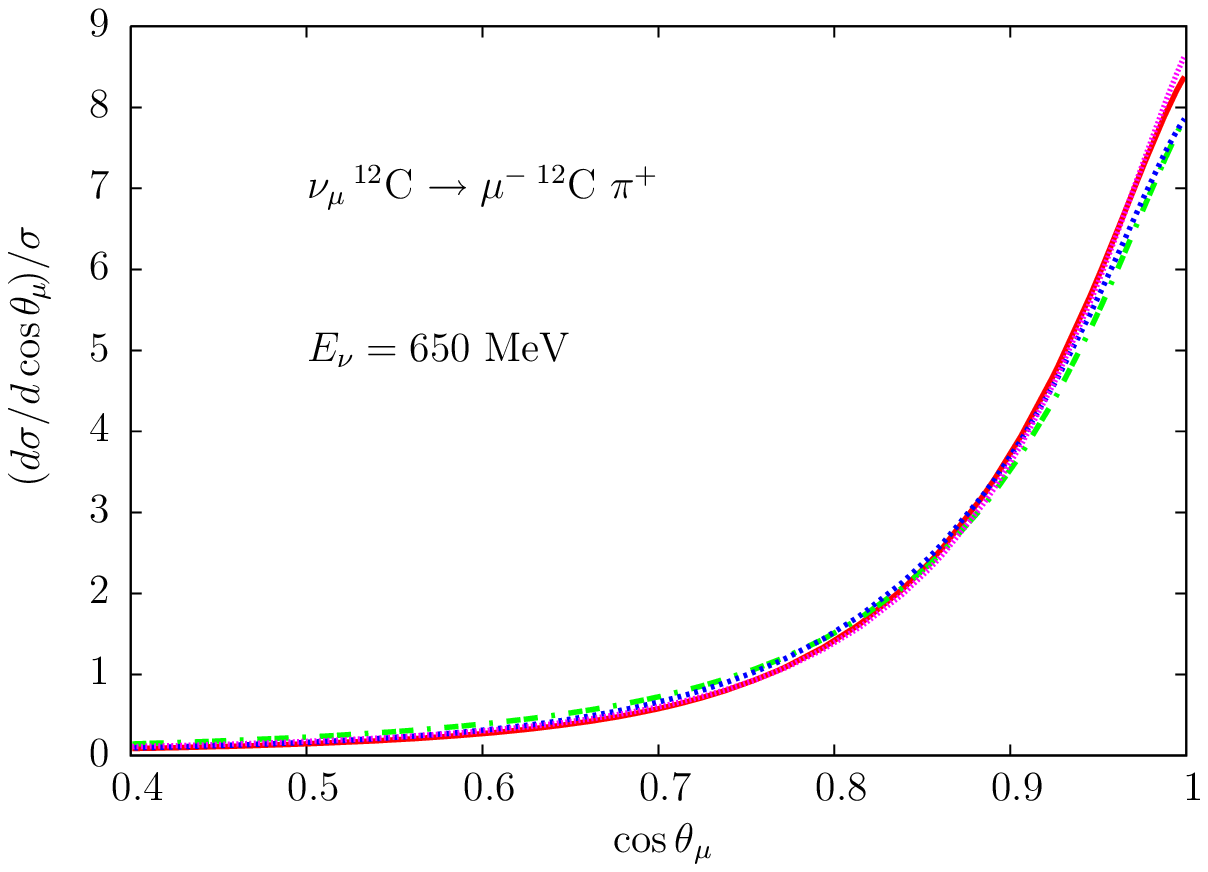}}
\end{center}
\caption{\footnotesize Muon angular  differential cross section for CC
  coherent pion production induced by a $650$ MeV energy $\nu_\mu$ beam on
  a $^{12}$C target. Left panel: absolute differential cross section;
  Right panel: differential cross sections normalized to one. The
  $\theta_\mu$ angle is referred to the incoming neutrino
  direction in the LAB frame.}\label{fig:dsdcosmu}
\end{figure}

We turn now to angular distributions of the final pion and muon in
$\nu_\mu$ induced CC reactions. In Fig.~\ref{fig:dsdcospi} we show the
pion angular LAB distribution with respect to the incoming neutrino
direction.  As expected, and due to the nucleus form factor, the
reaction is very forward peaked. Inclusion of medium effects on the
$\Delta$ propagator and the final pion distortion largely reduce the
cross section. The angular distribution profile keeps its forward
peaked behaviour, although less pronounced once the pion distortion is
included. This can be seen on the right panel where the angular
distributions are all equally normalized to one.  Such a behaviour can
be understood by taking into account that the pion wave function
$\widetilde{\varphi}_{\pi}^{\ast} (\vec{r};\vec{k}_\pi)$ has not well
defined momentum, in contrast with the plane wave $e^{-{\rm
i}\vec{k}_\pi\cdot\vec{r}}$. Putting the nucleons at rest has some
effect on the cross section but hardly affects the angular
distribution profile.
The situation is very similar for the muon angular distribution, shown in
Fig.~\ref{fig:dsdcosmu} for CC
coherent pion production by a $0.65$ GeV energy $\nu_\mu$ beam on a
$^{12}$C target. In this case the angular distribution profile is
completely unaffected by the nuclear effects on the $\Delta$ and the pion
distortion.

\subsection{The Rein--Sehgal Model and the NC MiniBooNE $E_\pi
  (1-\cos\theta_\pi)$ distribution}
\label{sec:rs}

\begin{figure}[htb]
\begin{center}
\makebox[0pt]{\includegraphics[scale=0.7]{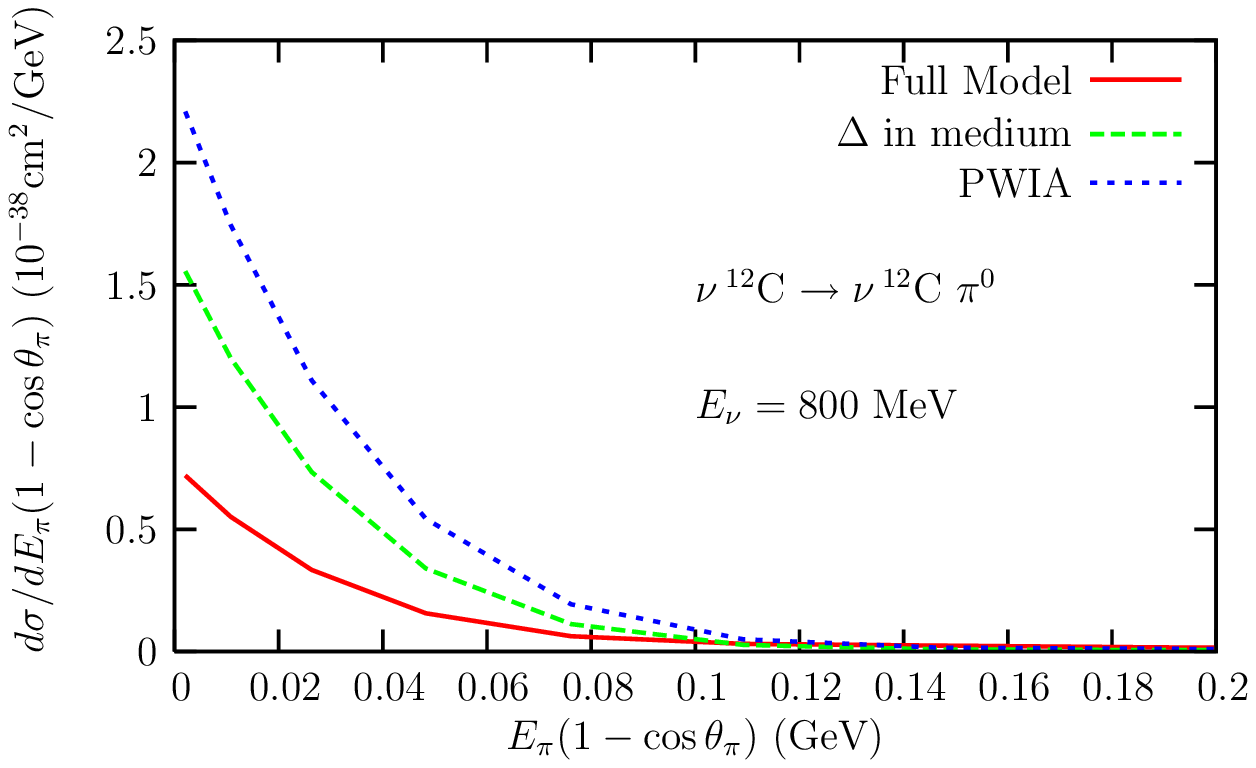}\hspace{0.5cm}
              \includegraphics[scale=0.7]{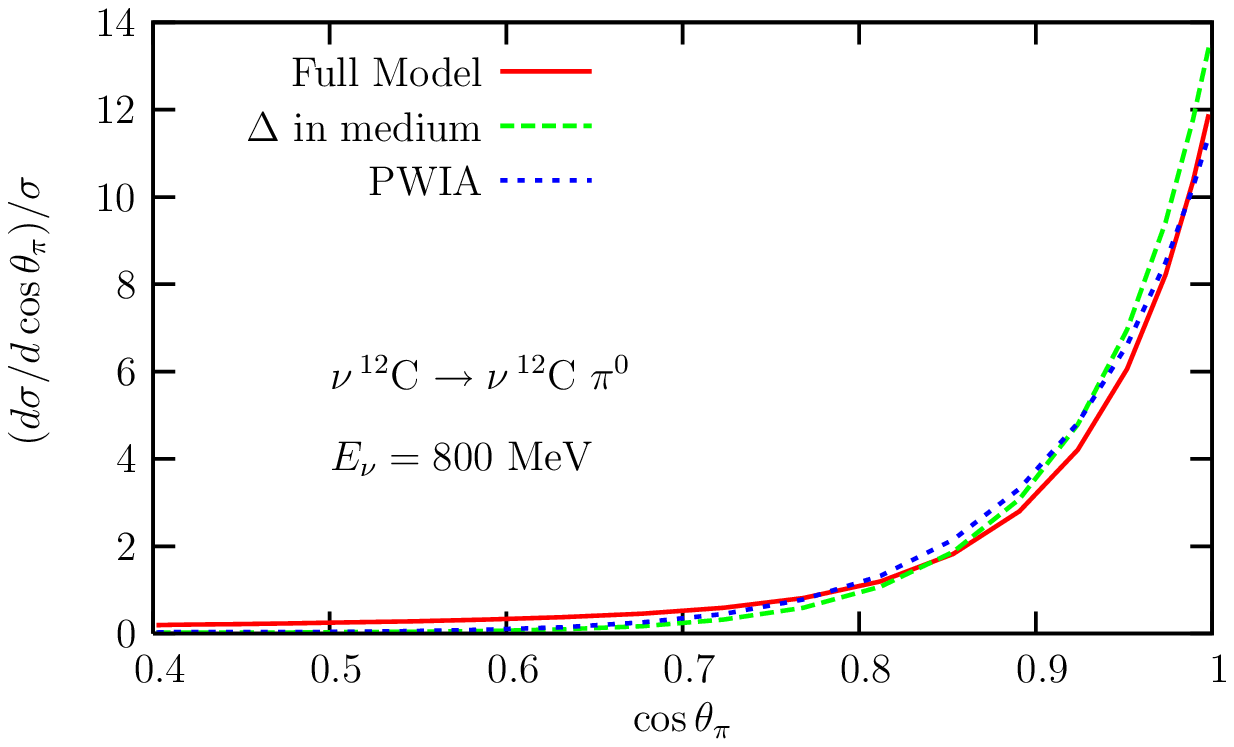}}\\\vspace{1.25cm}

\makebox[0pt]{\includegraphics[scale=0.35]{epicosteta.eps}\hspace{0.5cm}
              \includegraphics[scale=0.35]{miniboone_RS.eps}}

\end{center}
\caption{\footnotesize Laboratory $E_\pi (1-\cos\theta_\pi)$ and
  $\cos\theta_\pi$ distributions for the $\nu\, ^{12}{\rm C}\to \nu\,
  ^{12}{\rm C}\, \pi^0 $ reaction, at MiniBooNE energies. Different
  models are considered in the upper panels, while our full model is
  always used in the left bottom one, where we also show the $E_\pi
  (1-\cos\theta_\pi)$ distribution convoluted with the $\nu_\mu$ MiniBooNE flux
  (solid line). Details of the convolution are explained in the text
  (see Eq.~(\protect\ref{eq:conv}) and
  Table~\protect\ref{tab:res}). In the right bottom panel, we show
  results from the $C_5^A$ axial contribution of the $\Delta P$
  mechanism, neglecting pion distortion and $\Delta$ in medium
  effects (see the text for the
  explanation of the two curves). In both bottom panels, we
  display the MiniBooNE published histogram (solid), conveniently scaled down,
  taken from the right panel of Fig.3 in 
  Ref.~\protect\cite{AguilarArevalo:2008xs}. Finally, in the right
  bottom panel, we also show MiniBooNE results (dashed histogram)
  obtained by turning off the NUANCE FSI of the outgoing pion
  (G. Zeller private communication).  }\label{fig:fig_dxnc}
\end{figure}

We have also examined the NC differential cross section with respect
to the variable $E_\pi (1-\cos\theta_\pi)$, proposed by the MiniBooNE
Collaboration in its recent analysis of coherent $\pi^0$ production of
Ref.~\cite{AguilarArevalo:2008xs}. Pion variables are referred to the
LAB frame. Results are shown in the left top panel of
Fig.~\ref{fig:fig_dxnc} for a neutrino beam energy of 800 MeV (close
to the $\nu_\mu$ energy peak of the MiniBooNE experiment) on carbon. Our
differential cross section is appreciably narrower than that displayed
in Fig.~3b of Ref.~\cite{AguilarArevalo:2008xs}\footnote{Note however,
  that the $E_\pi (1-\cos\theta_\pi)$ distributions shown in
  Ref.~\protect\cite{AguilarArevalo:2008xs} are not proper
  differential cross sections. This is because, they have not been
  corrected for acceptance or cut efficiencies and are plotted for
  reconstructed kinematic quantities. So they include the effects of
  the selection criterion (the efficiency of which can vary as a
  function of $E_\pi(1-\cos\theta_\pi)$), as well as reconstruction
  effects in the MiniBooNE detector.  Currently, the MiniBooNE
  Collaboration is working to have all of the effects of the detector,
  event reconstruction, and selection removed. The new results will
  follow in an upcoming paper, where actual differential cross
  sections will be available~\cite{zeller}.}.  Indeed, our
distribution at $E_\pi (1-\cos\theta_\pi) = 0.1$ GeV has already
fallen off by a factor 18, while the MiniBooNE distribution has fallen
off by less than a factor 4 at 0.1 GeV, and even at 0.2 GeV the
reduction factor is still smaller than 15. We find hard to understand
this discrepancy. Since the excitation of $\Delta(1232)-$resonance
mechanism is dominant, we expect the outgoing pion to have a total
energy of around $0.25-0.35$ GeV (see for instance left bottom panel
of Fig.~\ref{fig:dsdppi}). On the other hand, we find that the cross
section is almost negligible for $\cos\theta_\pi > 0.7$, as can be
inferred from the pion angular LAB differential cross section
displayed in the right top panel of Fig.~\ref{fig:fig_dxnc}. Thus, we
easily understand why we find that the $E_\pi (1-\cos\theta_\pi) $
distribution becomes quite small above $0.1$ GeV. To have
non-negligible signals in the $E_\pi (1-\cos\theta_\pi)=0.1-0.2$ GeV
region would require, approximately, values of $\cos\theta_\pi$ in the
interval $2/3-1/3$, which would translate into values of $\theta_\pi $
in the range $50^\circ -70^\circ$. Since $\vec{q}$ is strongly aligned
to the incoming neutrino direction, such high pion angles look hardly
compatible with the forward character of the coherent reaction, which
is just imposed by the nucleus form factor (Fourier transform of the
nuclear density for momentum $\vec{q}-\vec{k}_\pi$) and the
$\vec{q}\cdot\vec{k}_\pi$ dependence of the amplitudes (dominated by
the axial contribution).

The strong disagreement between our prediction and the MiniBooNE
histogram in the region below $E_\pi (1-\cos\theta_\pi)$ < 0.1 GeV is
even most worrying, because there, the cross sections are much larger.
We are aware that within the $\nu_\mu$ MiniBooNE flux, there exist neutrino
energy components higher and smaller than the 800 MeV considered in
the top panels of Fig.~\ref{fig:fig_dxnc}.  In the left bottom panel
of Fig.~\ref{fig:fig_dxnc}, we show the $E_\pi (1-\cos\theta_\pi)$
differential cross section for three more neutrino energies ($300$,
$550$ and $1300$ MeV), in addition to that of 800 MeV considered in
the top panels. In all cases, we find very small signals for $E_\pi
(1-\cos\theta_\pi)$ above $0.05$ GeV. Also in this panel, we show our
$E_\pi (1-\cos\theta_\pi)$ differential cross section convoluted with
the $\nu_\mu$ MiniBooNE flux (solid line), and we certainly find a distribution
definitely narrower than that published in
Ref.~\cite{AguilarArevalo:2008xs}.  

Since the MiniBooNE analysis relies on the Rein--Sehgal model for
coherent $\pi^0$ production~\cite{Rein:1982pf}, the strong shape
difference should be understood in terms of the differences between
our model and that of Ref.~\cite{Rein:1982pf}.

\subsubsection{The $t-$dependence of the Rein--Sehgal model}
\label{sec:rs01}

Rein and Sehgal made use of the Adler's PCAC
formula~\cite{Adler:1964yx} and approximated (both for neutrino or
antineutrino induced processes) the coherent $\pi^0$ production
differential cross section by
\begin{equation}
\left( \frac{d\sigma_{\nu\nu}}{dxdy d|t|}\right)_{q^2=0} = \frac{G^2 M
E_\nu}{\pi^2}f_\pi^2 (1-y) \left (|F_{\cal A}(t)|^2 F_{\rm abs}
\frac{d\sigma(\pi^0 N \to \pi^0  N)}{d|t|}\Big|_{q^0=E_\pi, t=0}\right )
\label{eq:rs}
\end{equation}
with $x=-q^2/2Mq^0$, $y=q^0/E_\nu$, and
$t=(q-k_\pi)^2=-(\vec{q}-\vec{k}_\pi)^2$. Besides, the nuclear form
factor is calculated as $F_{\cal A}(t)=\int d^3\vec{r}\ e^{{\rm
i}\left(\vec{q}-\vec{k}_\pi\right)\cdot\vec{r}}
\left\{\rho_p(\vec{r}\,)+\rho_n(\vec{r}\,)\right \}$, and finally
$F_{\rm abs}$ is a $t-$independent attenuation factor\footnote{In the
original work of Ref.~\protect\cite{Rein:1982pf}, it is stated that
$F_{\rm abs}$ takes into account effects of pion absorption in the
nucleus. As defined in Ref.~\protect\cite{Rein:1982pf}, $F_{\rm abs}$
only removes from the flux pions that undergo inelastic collisions
but, as explained below, no true absorption is actually
included.}. The above expression was deduced in the so called parallel
configuration, for which the $k_\mu$ and $k^\prime_\mu$ four momenta
are proportional (therefore $q^2=0$) and $\cos\theta_{qk_\pi}$ (angle
formed by $\vec{k}_\pi$ and $\vec{q}\,$) and
$|\vec{k}_\pi|/|\vec{q}\,|$ are approximated to one everywhere except
in the nuclear form factor.  It was continued to non-zero $q^2$ values
by including a propagator term of the form $(1-q^2/m^2_A)^{-2}$, with
$m_A \approx 1$ GeV. The model should work well close to this parallel
kinematics, and constitutes a good approximation at the high neutrino
energies, above 2 GeV, explored in the original work of
Ref.~\cite{Rein:1982pf}.  However, the approximations in which is
based this model become less justified as the neutrino energy
decreases. For the energies relevant in the MiniBooNE and T2K
experiments, non parallel configurations turn out to be more
important, and the Rein--Sehgal model predictions are less reliable.

Actually, we see from Eq.~(\ref{eq:rs}) that the Rein--Sehgal
differential cross section depends on $\cos\theta_\pi$ or $t$ only
through the nuclear form factor, and any further $\cos\theta_\pi$
and/or $t$ behaviour induced by the dependence of the amplitudes on
$k_\pi$ is totally neglected.  However, it is reasonable to expect
that these additional angular dependences might play a role when the
pion emission is not completely forward. To illustrate this point, we
have re-derived Eq.~(\ref{eq:rs}) from our model. We have considered
the dominant $C_5^A$ axial contribution of the $\Delta P$ mechanism
and made use, in this case, of the Goldberger--Treiman relation to express $C_5^A(0)$
in terms of the $\pi N \Delta$ coupling, $f^*$, and the pion decay
constant $f_\pi$ $\left [C_5^A(0) =
\sqrt\frac23\frac{f_\pi}{m_\pi}f^*\right]$.  Besides, we have
considered neither the in medium $\Delta-$selfenergy, nor the pion
distortion effects.  On the other hand, we have used the $N\Delta\pi$
Lagrangian\footnote{Here, $\vec{\phi}$ is the pion field, $\Psi_\mu$
is a Rarita Schwinger $J^\pi = 3/2^+$ field, $\vec{T}^\dagger$ is the
isospin transition operator (vector under isospin rotations and its
Wigner-Eckart irreducible matrix element is taken to be one) from
isospin 1/2 to 3/2.}
\begin{equation}
{\cal L}_{\pi N \Delta} = \frac{f^*}{m_\pi} \bar\Psi_\mu
\vec{T}^\dagger (\partial^\mu \vec{\phi}) \Psi + {\rm h.c.}
\label{eq:lagranpind}
\end{equation}
to compute $d\sigma(\pi^0 N \to \pi^0 N)/d|t|$. To further simplify,
we have worked in the non-relativistic limit for the baryons. Apart from $F_{\rm
abs}$, we
reproduce Eq.~(\ref{eq:rs}), but with an extra
$(\cos\theta_{qk_\pi})^2$ factor in the right hand side of the equation,
which for the case of parallel kinematics is equivalent to $(\cos
\theta_\pi)^2$. At high neutrino energies, the pion is emitted
strongly forward and thus, it is consistent to approximate this factor
by one, as it was done in the original work of Ref.~\cite{Rein:1982pf}
to obtain Eq.~(\ref{eq:rs}). For MiniBooNE
neutrino energies, relatively large $\theta_\pi$ angles are allowed,
in special for low pion momenta ($|\vec{k}_\pi| < 0.2$
GeV)~\cite{Link}, and this factor $(\cos \theta_\pi)^2$ could make the
$E_\pi (1-\cos\theta_\pi)$ distribution significantly narrower than
that predicted by the Rein--Sehgal model. For a fixed value of $E_\pi
(1-\cos\theta_\pi)$, the effect becomes more important as the pion
energy decreases.  For instance, if we fix $E_\pi
(1-\cos\theta_\pi)=0.05$ GeV, we find that $(\cos\theta_\pi)^2 \approx
1/2 $ or 3/4, for an averaged pion energy of 0.17 or 0.375 GeV,
respectively. As it was shown in the talk of J. Link at
Nuint07~\cite{Link}, in the Rein--Sehgal model, low energy pions
($|\vec{k}_\pi| < 0.5$ GeV) produce wider $E_\pi (1-\cos\theta_\pi)$
shapes than those of higher energies ($|\vec{k}_\pi| > 0.5$ GeV), and
thus we expect this type of corrections to be important. In addition,
there also exist corrections that vanish in the $q^2 \to 0$ (or
equivalently $|\vec{q}\,|/q^0 \to 1$ ) and/or
$|\vec{k}_\pi|/|\vec{q}\,|\to 1$ (also implied by the $t=0$
approximation in the amplitudes) limits.

Another way to see the limitations of the Rein--Sehgal model is the
following. As mentioned, this model assumes no further dependence on
$t$ than that encoded in the nuclear form factor. Since $t=0$ implies
$q=k_\pi$, we have replaced $k^\alpha_\pi$ in
Eq.~(\ref{eq:prescription}) by $q^\alpha$. It is to say, we replace
$k_\pi$ by $q$ in the pion emission vertex\footnote{ If we were to repeat with
this replacement the derivation of Eq.~(\ref{eq:rs}) as explained
above, we will recover it exactly 
(apart from $F_{\rm abs}$), and the correction factor
$(\cos\theta_{qk_\pi})^2$ would not appear.}.  To better compare with
the Rein--Sehgal predictions, we have again just considered the
dominant $C_5^A$ axial contribution of the $\Delta P$ mechanism,
without considering pion distortion and $\Delta$ in medium
effects. $\nu_\mu$ MiniBooNE flux convoluted results   are displayed
in the right bottom panel of Fig.~\ref{fig:fig_dxnc}. We see that the
new $E_\pi(1-\cos\theta_\pi)$ distribution is significantly wider than
that obtained without implementing this replacement, and that it
reasonably describes the MiniBooNE published distribution (solid
histogram in the plot). The agreement is much better, when we
compare with some preliminary MiniBooNE results (dashed histogram)
obtained with a different treatment of the outgoing pion Final State
Interaction (FSI), as we will explain in the next
sub-subsection\footnote{We are indebted with G. Zeller for providing
us these preliminary results.}. Without giving a special meaning to this 
agreement, this simple calculation serves the purpose of 
illustrating the uncertainties associated to the $t=0$ approximation
at low energies, for which the nuclear form factor still allows some
deviations from the completely forward scattering.

We conclude that, the Rein--Sehgal pion coherent production model for
MiniBooNE and T2K experiments is not as reliable as for the case of
neutrino energies above 2 GeV. We expect sizable corrections to the
predictions of this model, both for differential distributions and for
integrated cross sections. Our model provides an $E_\pi
(1-\cos\theta_\pi)$ distribution much more peaked around zero, and
thus it might improve the description of the first bin value in
Fig.~3b of Ref.~\cite{AguilarArevalo:2008xs}. Moreover, the drastic
change in the $E_\pi (1-\cos\theta_\pi)$ distribution shape might
produce some mismatch between the absolute normalization of the
background, coherent and incoherent yields in the MiniBooNE analysis.
One the other hand, and besides of the issue of the used
value\footnote{Within the Rein--Sehgal model, this constant is
implicitly fixed to approximately 1.2, since the Goldberger--Treiman
relation is used to express the coherent $\pi^0$ production cross
section in terms of the elastic $\pi^0 N \to \pi^0 N$ one.} for
$C_5^A(0)$, the Rein--Sehgal model, adopted by the MiniBooNE
Collaboration in Ref.~\cite{AguilarArevalo:2008xs}, overestimates by a
large factor the coherent integrated cross section for MiniBooNE energies.
This is due first to the $t=0$ approximation in the amplitudes assumed
in this model that produces a too wide $E_\pi (1-\cos\theta_\pi)$
distribution, which leads to cross sections larger by about a factor
of two than those obtained when the $t-$dependence is properly taken
into account (see for instance the different areas below the solid and
dashed-dotted curves in the right bottom panel of
Fig.~\ref{fig:fig_dxnc}). Secondly, because all sort of
in-nuclear-medium effects, like pion absorption or modification of the
elementary $\pi N \to \pi N$ cross section\footnote{Within our model,
we include these modifications by means of the consideration of the
$\Delta-$selfenergy, by considering the Pauli blocking in the
computation of $\Delta-$decay width in the medium, and by using the
pion wave function, $\widetilde{\varphi}_{\pi}^{\ast}$, instead of a plane
wave.} inside of the nucleus..., which are not accounted for in the
Rein-Sehgal model, and that turn out to be relevant at MiniBooNE
energies.

\subsubsection{The $E_\pi (1-\cos\theta_\pi)$ > 0.1 GeV region and NUANCE FSI}

We turn now to the region $E_\pi (1-\cos\theta_\pi)$ > 0.1 GeV. Also
 in this region, the bulk of the discrepancies among our predictions
 and the MiniBooNE results can be understood in terms of the
 inaccuracies of the Rein--Sehgal model at low energies, as can be
 appreciated in the right bottom panel of Fig.~\ref{fig:fig_dxnc}. But
 here, the outgoing $\pi^0$ FSI effects, incorporated by the analysis
 of Ref.~\cite{AguilarArevalo:2008xs}, induce some additional
 discrepancies.  The MiniBooNE analysis relies on the Rein--Sehgal
 models for incoherent (resonant)~\cite{Rein:1980wg} and
 coherent~\cite{Rein:1982pf} $\pi^0$ production, which are implemented
 in the NUANCE event generator~\cite{Hawker:2005vf}.  In the case of
 coherent production, the Rein--Sehgal model includes an absorption
 factor $F_{\rm abs}$ to effectively account for the pion wave
 function distortion from a plane wave (see Eq.~(\ref{eq:rs})). This
 Glauber-type factor decreases exponentially with the pion-nucleon
 inelastic cross section (see Eq. (24) of Ref.~\cite{Rein:1982pf}) and
 thus it neither accounts for pion absorption, which is a two nucleon
 mechanism, nor it does for quasielastic distortion.  Quasielastic
 steps, induced by elastic pion-nucleon collisions and not forbidden
 by Pauli blocking, excite and/or break the nucleus, and are not
 removed by this factor $F_{\rm abs}$. This Glauber factor only
 removes events where the $\pi^0$ suffers an inelastic collision with
 a nucleon, and as a result changes its charge, or other mesons are
 produced in the final state.

 The procedure followed in Ref.~\cite{AguilarArevalo:2008xs} to
 describe coherent pion production is somewhat different. They set
 $F_{\rm abs}$ to 1 in the Rein--Sehgal model of
 Ref.~\cite{Rein:1982pf}, and implement absorption as part of the FSI.
 In our understanding, coherent pion production cross section cannot
 be calculated from a Monte Carlo cascade algorithm. This is because,
 by definition, the coherent production is a one step
 process\footnote{The nomenclature here might be confusing.  There
 exist multiple step contributions to the coherent reaction. For
 instance, a $\Delta$ is formed in a NC scattering, it decays with the
 nucleon falling back into the hole created by $\Delta$ formation, the
 decay $\pi^0$ creates a subsequent $\Delta$, which in turn decays
 emitting a $\pi^0$ that escapes the nucleus and the associated
 nucleon also drops into the ground state configuration. The point we
 want to make here is that such contributions cannot be taken into
 account in a Monte Carlo cascade algorithm. This is because it would
 require the coherent sum of the multiple step amplitudes, while a Monte Carlo
 alogorithm uses probabilities (cross sections). We do include these
 multiple step contributions within our formalism thanks to the use of
 a pion wave function solution of the Klein-Gordon equation,  with an
 optical $\pi^0$-nucleus potential, instead of using a plane wave.},
 and the quantum mechanical transition matrix element gives the
 amplitude probability for producing a pion outside of the nucleus,
 which is left unchanged. The coherent contribution should be
 incoherently added to that due to the inelastic channels to find out
 the total pion production cross section.

 Nevertheless, one could still reasonably estimate the total coherent
 cross section from the NUANCE FSI cascade, if it would be used to
 eliminate from the flux of outgoing neutral pions, not only those
 which get absorbed or those that suffer inelastic processes, like
 multiple pion production, meson production or pion charge exchange,
 but also those that undergo quasielastic steps, induced by elastic
 pion-nucleon collisions, in their
 way out of the nucleus. However, to our knowledge, these latter
 events are accounted for in the MiniBooNE analysis, despite they are
 not coherent since the final nucleus, as a result of the secondary
 collisions, is not left in the ground state.  We believe this is
 acknowledged by the authors of Ref.~\cite{AguilarArevalo:2008xs} when
 they say ``{\it ...that rescattered events with a $\pi^0$ in the
 final state may be misclassified in NUANCE, as would be the case when
 a coherently produced $\pi^0$ rescatters elastically through a
 resonance}''.  As a result of these collisions, the $\pi^0$'s might
 change their direction and give rise to events in the NUANCE cascade
 at significantly larger values of $\theta_\pi$.  Indeed in the right
 bottom panel of Fig.~\ref{fig:fig_dxnc}, we observe significantly
 less events above $E_\pi (1-\cos\theta_\pi)$ > 0.1 GeV when the
 NUANCE FSI is turned off (dashed histogram)\footnote{Note that when
 NUANCE FSI is turned off, besides of getting rid of the unwanted
 quasielastic steps, pion absorption is not taken into account.
 However, this latter effect though, produces a  diminution
 of events, it does not significantly change the shape of the
 $E_\pi(1-\cos\theta_\pi)$ distribution. On the other hand, since the
 histogram has been re-scaled down, the overall normalization is not
 an issue any more.  Nevertheless, we should point out that, there
 could be same minor differences in the acceptance or cut efficiencies
 with respect those used in the published
 histogram~\cite{AguilarArevalo:2008xs}.}.  The effects below 0.1 GeV
 are much smaller, and in total, the change in the shape leads to a
 reduction of around 20\% in the integrated cross section.

In our calculation of the coherent cross
section, we certainly remove those secondary events by means of the optical
potential employed to compute the pion wave function. The imaginary
part of the pion-nucleus potential is responsible for the removal of
flux of the outgoing pions on their way out of the nucleus. This
imaginary part is due to pion absorption,  but also to pion quasielastic
steps.  Hence, the use of the full optical potential will eliminate
the pions which are absorbed and also those which scatter
quasielastically. We might try to theoretically estimate this
effect by switching off the quasielastic contribution to the
pion-nucleus optical potential induced by elastic pion-nucleon
collisions, and using an optical potential with an imaginary part due
to absorption and inelastic channels alone. In this way, we will
remove the absorbed pions and those that undergo inelastic collisions,
but not those which scatter quasielastically, which will still go out
of the nucleus and are accounted for by the MiniBooNE Monte Carlo
generator. This was considered in the past in the study of the pionic
decay of $\Lambda-$hypernuclei~\cite{Nieves:1992pm,Straub:1992yw},
leading to moderate enhancements of the decay widths of the order of
$10-15\% $ in $^{12}$C~\cite{Albertus:2002kk}. We find here similar
effects, and for the MiniBooNE flux averaged cross section, we find
an enhancement of around 20\% (see Table~\ref{tab:res}), 
in good agreement with the effects observed by turning off the NUANCE FSI.

\subsection{K2K, MiniBooNE and T2K Flux Averaged Cross Sections}

\begin{figure}[htb]
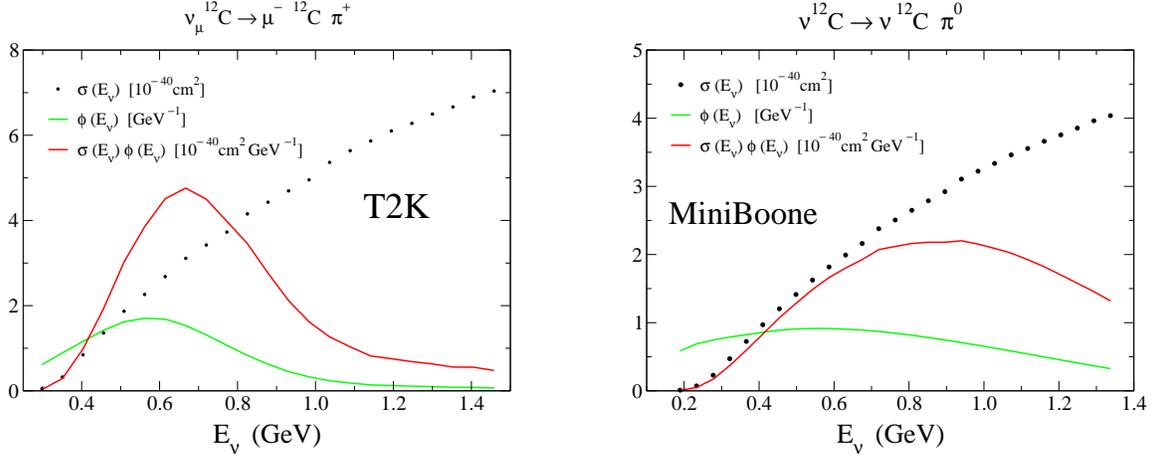

\begin{center}
\makebox[0pt]{\includegraphics[scale=0.3]{cct2k_v2.eps}\hspace{1.5cm}
              \includegraphics[scale=0.3]{ncminiboone_v2.eps}}\\
\end{center}
\caption{\footnotesize CC (left) and NC (right) coherent pion
 production cross sections in carbon. We also show predictions
 multiplied by the T2K (left) and MiniBooNE (right) $\nu_\mu$ neutrino energy
 spectra.  In the region of neutrino energies around 0.6 GeV, the
 lower curves stand for the T2K and MiniBooNE $\nu_\mu$ fluxes normalized to
 one. \\ \vspace{0.5cm}}\label{fig:flux-cross}
\end{figure}

In Table~\ref{tab:res} we show our predictions for the
K2K~\cite{Hasegawa:2005td} and MiniBooNE~\cite{AguilarArevalo:2008xs}
flux averaged cross sections as well as for the future T2K
experiment. In Fig.~\ref{fig:flux-cross}, we show some results for K2K
and MiniBooNE experiments. In all cases, we normalize the neutrino or
antineutrino flux $\phi$ to one. In principle, we would like to
compute the corresponding convolution with the neutrino or
antineutrino flux
\begin{equation}
\bar \sigma =\int_{E_{\rm low}^i}^{E_{\rm high}^i} dE \phi^i(E)
\sigma(E), \quad i ={\rm K2K,~MiniBooNE,~T2K}
\end{equation}
with $E_{\rm low}^i, E_{\rm high}^i$ the lower and upper flux limits,
and $\sigma(E)$ the corresponding CC/NC muon/electron
neutrino/antineutrino induced nuclear coherent cross section, as a
function of the neutrino/antineutrino energy\footnote{Note that the
cross section trivially vanishes for neutrino/antineutrino energies
below the pion production threshold, which obviously is different for
CC and NC driven processes because of the final lepton mass.}. In
practice, the predictions of our model become less reliable
when the energy increases, since the model neglects all resonances
above the $\Delta(1232)$. Sophisticated recent calculations, as those of 
Refs.~\cite{Singh:2006bm, AlvarezRuso:2007tt, AlvarezRuso:2007it}, 
suffer from exactly the same limitation.  That is the reason why we
have set up a maximum neutrino energy ($E_{\rm max}^i$) in the convolution, and
approximated
\begin{equation}
\bar \sigma \approx \frac{\int_{E_{\rm low}^i}^{E_{\rm max}^i} dE \phi^i(E)
\sigma(E)}{\int_{E_{\rm low}^i}^{E_{\rm max}^i} dE \phi^i(E)} \label{eq:conv}
\end{equation}
where we fix the upper limit in the integration (neglecting the long
tail of the neutrino fluxes) to $E_{\rm max}=1.45$ and 1.34 GeV for CC
and NC muon neutrino/antineutrino driven processes, respectively.  The
phase space for the fifth differential
$\frac{d^{\,5}\sigma}{d\Omega(\hat{k^\prime})dE^\prime
d\Omega(\hat{k}_\pi) }$ cross section, up to irrelevant constants, is
determined by $|\vec{k}'| |\vec{k}_\pi|$ [see Eqs~(\ref{eq:sec}),
(\ref{eq:Jmunu}) and (\ref{eq:nc-sec}), (\ref{eq:zmunu})]. For CC
processes, for instance, and muon neutrino energies of around 1.45
GeV, the phase space peaks at pion energies of around 730 MeV, which
leads, neglecting the nucleon momentum, to $\pi N$ invariant masses
below 1.5 GeV. Up to these energies, one can reasonably assume $\Delta
(1232)$ dominance.

In the case of the K2K experiment a threshold of 450 MeV for muon
momentum was imposed as an additional selection
criterion~\cite{Hasegawa:2005td}. We have implemented this cut also
here, and in that case we have been able to go up to $E_{\rm max}^{\rm
CC,K2K}$=1.8 GeV. In these circumstances, we still cover about 90\% of
the total flux in most of the cases. For the T2K antineutrino flux, we
cover just about 65\% of the total spectrum, and therefore our results
for the convoluted cross sections are less reliable.

For neutrino energies above 1 GeV, and though the $\Delta$
contribution  plays a central role in pion
production~\cite{Rein:1980wg}, one should bear in mind that other
resonances could certainly be also important. This would affect the
results presented in Table~\ref{tab:res} and
Fig.~\ref{fig:flux-cross}. Taking into account that the T2K and
MiniBooNE fluxes peak at neutrino energies of around 0.6--0.7 GeV, where
the $\Delta$ resonance contribution is much more dominant, is
reasonable to expect corrections (higher cross sections) of around
20-30\% to our results for these two experiments. Certainly, the
corrections could be larger for the K2K case, since it that case the
neutrino energy spectrum peaks at higher energies, around 1.2 GeV. 

We see that our prediction, subject to some uncertainties, lies well
below the K2K upper bound, mainly thanks to the use of a low value for
$C_5^A(0)$, while we predict a NC MiniBooNE cross section notably
smaller than that given in the PhD thesis of J.L. Raaf~\cite{Raaf}.
However, this latter value should be taken with extreme caution. It
was obtained from a preliminary analysis, that since then has been
notably improved. Moreover, the MiniBooNE Collaboration has not given
an official value for the total coherent cross section yet, and only
the ratio coherent/(coherent+incoherent) has been
presented~\cite{AguilarArevalo:2008xs}.  Nevertheless, as we have
discussed at length, we believe the MiniBooNE analysis might
overestimate this ratio, not only because some of the $\pi^0$'s which
undergo FSI collisions are accounted for as coherent events instead of
being removed, but more importantly because the Rein--Sehgal model
predicts an incorrect (wider) $E_\pi (1-\cos\theta_\pi)$ shape for
coherent $\pi^0$'s. The first of the effects produces an enhancement
of the coherent cross section of the order of 20\% (see the MiniBooNE
NC* entry in the table), while it is much more difficult to quantify
the second of the effects. This is because, in addition to the
variation of the integrated area, it might produce a possible
mismatch between the absolute normalization of the background,
coherent and incoherent yields in the MiniBooNE analysis.  To finish
this discussion, we would like to point out that the K2K cross section
and the value quoted in Ref.~\cite{Raaf} seems somehow incompatible
with the approximate relation $\sigma_{\rm CC} \approx 2 \sigma_{\rm
  NC}$, expected from $\Delta-$dominance and neglecting finite muon
mass effects (see discussion at the end of Sect.~\ref{sec:nc}).

Our predictions are about 20-30\% smaller than those obtained in
Ref.~\cite{AlvarezRuso:2007it} from model II, 
which uses  our value $C_5^A(0)\approx
0.9$. This discrepancy is due to different facts: the existence
of some numerical inaccuracies affecting the results of
Refs.~\cite{AlvarezRuso:2007tt,AlvarezRuso:2007it},  the
inclusion of the non zero momenta in the  nucleon spinors, different
ranges in the flux convolution, etc...

For the future T2K experiment, we get cross sections of the order
2$-$2.6$\times10^{-40}$cm$^2$ in carbon and about 2.3$-$3.0 $\times
10^{-40}$cm$^2$ in oxygen, considerably smaller than those predicted
by the Rein--Sehgal model.

The SciBooNE Collaboration has recently set 90\% confidence level
upper limits on the cross section ratio of CC coherent pion production
to the total CC cross section at $0.67 \times 10^{-2}$ at mean
neutrino energy 1.1 GeV and $1.36 \times 10^{-2}$ at mean neutrino
energy 2.2 GeV~\cite{Hi08}.  If we use a value of $1.05 \times
10^{-38}$ cm$^2$/nucleon for the total CC cross section, as quoted
in~\protect\cite{Hi08}, the SciBooNE upper limit for the ratio will
transform in a upper bound of about $8\times 10^{-40}$ cm$^2$ for the
coherent cross section at 1.1 GeV. At 1.1 GeV, our prediction for the
CC coherent pion production cross section in carbon is $5.7 \times
10^{-40}$ cm$^2$ (see points in the left panel of
Fig.~\ref{fig:flux-cross}), which is totally compatible with the
SciBooNE bound. However, according to the Rein-Sehgal
model~\cite{Rein:1982pf,Rein:2006di} implemented in the SciBooNE
simulation, the cross section ratio of CC coherent pion production to
the total CC cross section is calculated to be $2.04\times 10^{-2}$ in
Ref.~\cite{Hi08}. The SciBooNE limits correspond to 33\% and 67\% of
the Rein-Sehgal model prediction at 1.1 GeV and 2.2 GeV,
respectively~\cite{Hi08}. From our discussion in Subsect.~\ref{sec:rs}
(see last paragraph in the sub-subsection~\ref{sec:rs01}), we easily
understand why the Rein-Sehgal model overestimates the coherent cross
sections by a large factor at neutrino energies around 1 GeV
(approximately a factor of two because of the value\footnote{Note,
  however that, $C_5^A(0)$ will partially cancel out in the ratios
  measured by the SciBooNE Collaboration.}  of $C_5^A(0)$, and
approximately another factor of two because of the $t=0$ approximation
in the amplitudes, in addition of all sort of in-nuclear-medium
effects, like pion absorption or modification of the elementary $\pi N
\to \pi N$ cross section in the medium,...  which are not accounted
for in the Rein-Sehgal model). We also understand why at the higher
neutrino energy of 2.2 GeV the Rein-Sehgal model works better, since
the larger the energy, the better the $t=0$ approximation in the
amplitudes and smaller the nuclear effects become.  Note also, that at
2.2 GeV, we expect heavier resonances than the $\Delta(1232)$ to play
an important role, and thus the issue of the value of $C_5^A(0)$ is
less relevant.

\begin{figure}[htb]
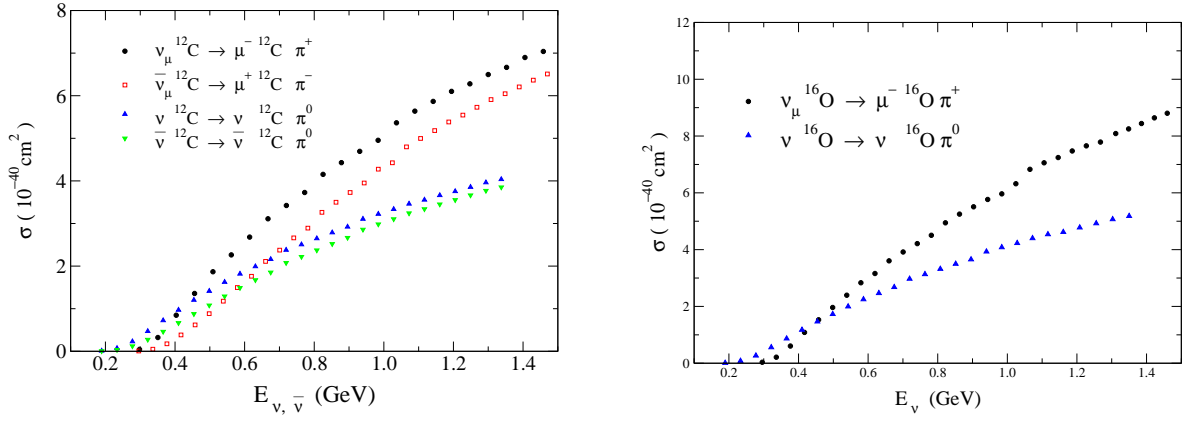

\begin{center}
\makebox[0pt]{\includegraphics[scale=0.3]{secefcarbono.eps}\hspace{1.cm}
              \includegraphics[scale=0.3]{secefoxigeno.eps}}\\
\end{center}
\caption{\footnotesize Muon neutrino/antineutrino CC
and NC coherent pion production off nuclei from carbon (left) and
oxygen (right) targets as a function of the neutrino/antineutrino energy. }\label{fig:cross}
\end{figure}

To conclude this section, in Fig.~\ref{fig:cross} we show muon
neutrino/antineutrino CC and NC coherent pion production off carbon
and oxygen targets. We see that both for CC and NC driven processes,
the ratio of neutrino over antineutrino cross sections approaches one
as the neutrino energy increases. This is due to the nuclear form
factor which reduces the vector contribution to the amplitudes (and
therefore the interference between the vector and the axial parts) as
the neutrino energy increases. Besides, and as a consequence of the
nuclear form factor and other nuclear in medium effects, we also see
that cross sections do not scale as $A^2$, being $A$ the nuclear mass
number, as expected from a coherent
reaction. This is in good agreement with the findings of
Refs.~\cite{Rein:1982pf,AlvarezRuso:2007tt}.  Indeed at 1 GeV, oxygen
and carbon cross sections turn out to be in a proportion of around 6
to 5, instead of 1.8 to 1, as it would be deduced from an $A^2-$type
scaling law. Finally, we observe sizable corrections to the
approximate relation $\sigma_{\rm CC} \approx 2 \sigma_{\rm NC}$ for
these two isoscalar nuclei in the whole range of neutrino/antineutrino
energies examined in this work. As pointed out in
Refs.~\cite{Rein:2006di,Berger:2007rq}, this is greatly due to the
finite muon mass, and thus the deviations are dramatic at low neutrino
energies. In any case, these corrections can not account for the 
apparent incompatibility among  the CC K2K cross section and the NC
value quoted in Ref.~\cite{Raaf}, mentioned above.

\begin{table}\begin{center}
    \begin{tabular}{lcccccc}\hline\tstrut
Reaction                 & Experiment &$\bar \sigma$& $\sigma_{\rm exp}$ &
$E_{\rm max}^i $ ~~ & $\int_{E_{\rm low}^i}^{E_{\rm max}^i} dE \phi^i(E)
\sigma(E)$~~ & $\int_{E_{\rm low}^i}^{E_{\rm max}^i} dE \phi^i(E)$
\\
&  & [$10^{-40}$cm$^2$] & [$10^{-40}$cm$^2$] & [GeV]
&[$10^{-40}$cm$^2$]  \\\hline\tstrut
CC\phantom{*} $\nu_\mu + ^{12}$C    & K2K        &   4.68    &
$<7.7 $~\cite{Hasegawa:2005td}   &1.80&3.84&0.82         \\
CC\phantom{*} $\nu_\mu + ^{12}$C    & MiniBooNE  &   2.99    &          &1.45&2.78&0.93 \\

CC\phantom{*} $\nu_\mu + ^{12}$C    & T2K        &   2.57    &          &1.45&2.34&0.91           \\
CC\phantom{*} $\nu_\mu + ^{16}$O    & T2K        &   3.03    &    &
  1.45& 2.76&  0.91           \\
NC\phantom{*} $\nu_{{\mu}} + ^{12}$C    & MiniBooNE  & 1.97 & $7.7\pm1.6\pm3.6$~\cite{Raaf}
&1.34&1.75&0.89\\
NC* $\nu_{{\mu}} + ^{12}$C    & MiniBooNE  & 2.38$^*$ &
$7.7\pm1.6\pm3.6$~\cite{Raaf} 
&1.34&2.12$^*$&0.89\\
NC\phantom{*} $\nu_{{\mu}} + ^{12}$C    & T2K        & 1.82     &     &1.34 &1.64
&0.90 \\
NC\phantom{*} $\nu_{{\mu}} + ^{16}$O    & T2K        & 
  2.27      & &  1.35     &  2.04 &  0.90  \\
CC\phantom{*} $\bar\nu_\mu + ^{12}$C    & T2K        &   2.12 &
&    1.45      & 1.42 & 0.67          \\
NC\phantom{*} $\bar\nu_{{\mu}} + ^{12}$C    & T2K        &
 1.50     &     &1.34 &0.96
&0.64 \\
\hline
    \end{tabular}
  \end{center} 
  \caption{\footnotesize NC/CC muon neutrino and
    antineutrino coherent pion production total cross sections for
    K2K, MiniBooNE and T2K experiments. In the
    case of CC K2K, the experimental threshold for the muon momentum
    $|\vec{k}^\prime|$ > 450 MeV is taken into account. To convert the cross
    section ratio given in~\protect\cite{Hasegawa:2005td} into a
    coherent cross section (K2K), we use the value of
    $1.07 \times 10^{-38}$ cm$^2$/nucleon for the total CC cross
    section, as quoted in~\protect\cite{Hasegawa:2005td}. For the MiniBooNE
    NC* entry, we present our results when an optical pion-nucleus
    potential with an imaginary part due to absorption and inelastic channels
    alone is used to compute the distortion of the outgoing pion (see
    text for more details). The absolute NC $\pi^0$ coherent cross
    section quoted in the PhD thesis of Ref.~\protect\cite{Raaf} should be
    taken with extreme caution, since in the  published paper
    (Ref.~\protect\cite{AguilarArevalo:2008xs}) it is not
    given. There, it is  quoted  the ratio of the sum of the
    NC coherent and diffractive modes over all exclusive NC $\pi^0$
    production at MiniBooNE.  Some details on the
    flux convolution are compiled in the last three columns. }
\label{tab:res} 
\end{table}

\section{Conclusions}

We have developed a model for neutrino/antineutrino CC and NC coherent
pion production off nuclei which is based on a microscopic model for
neutrino/antineutrino-induced one-pion production off the nucleon
derived in Ref.~\cite{Hernandez:2007qq}. The model of
Ref.~\cite{Hernandez:2007qq} includes the dominant $\Delta$ production
mechanism, but it also takes into account background terms required by
chiral symmetry. In its application to coherent production we have
further taken into account the main nuclear effects expected to be
important in reactions off nuclei.  While the model presented here is
similar to the one in
Refs.~\cite{AlvarezRuso:2007tt,AlvarezRuso:2007it}, we have improved
on that calculation by taking consistently into account the nucleon
motion, and by using a more sophisticated pion optical potential. The
consideration of the nucleon motion increases the cross section by a
non negligible amount, while Coulomb effects on the emission of
charged pions lead to small changes in the cross section. Moreover, we
have corrected for some numerical inaccuracies~\cite{luis} (of the
order of 20\% in the total cross section, and larger at the peak of
the $d\sigma/dk_\pi$ differential distribution) that affected the
calculations carried out in
Refs.~\cite{AlvarezRuso:2007tt,AlvarezRuso:2007it}.

In agreement with Refs.~\cite{AlvarezRuso:2007tt,AlvarezRuso:2007it},
we find a strong reduction of the cross section, mainly due to the
modification of the $\Delta$ self-energy in the nuclear medium, and a
shift to lower energies of the outgoing pion distribution due to the
final pion distortion. The angular distributions of both pion and
muons with respect to the incoming $\nu_\mu$ direction are forward
peaked due to the nuclear form factor. While the muon angular
distribution profile is almost unaffected by nuclear corrections, in
the pion case, part of the strength is shifted to larger angles due to
the distortion of the final pion wave function.  Non-resonant terms,
that turned out to be very important at the nucleon
level~\cite{Hernandez:2007qq}\footnote{Their inclusion made necessary
to reduce the nucleon-to-$\Delta$ resonance axial coupling.}, give
small contributions to the coherent pion production off isospin
symmetric nuclei.  This leads us to find coherent pion production
cross sections around a factor of two smaller than most of those
previously published.

We have also performed a detailed discussion of the MiniBooNE results
of Ref.~\cite{AguilarArevalo:2008xs} and the analysis performed there
to identify NC coherent $\pi^0$ events. We have shown that the
Rein--Sehgal model used in this analysis is not accurate enough in
this case. This is because the MiniBooNE flux mainly consists of
neutrinos below 2 GeV, and for such low neutrino energies, the
corrections to the outgoing pion angular dependence predicted by the
Rein--Sehgal model become quite important. As a consequence, the
Rein--Sehgal model leads to distributions notably wider and integrated
cross sections much larger than those
predicted in this work. Finally, we have predicted muon
neutrino/antineutrino CC and NC coherent pion production off carbon
and oxygen up to neutrino energies of the order of 1.4 GeV, and
convoluted those cross sections with the K2K, T2K and MiniBooNE
fluxes. Our cross sections are considerably smaller than those
predicted by the Rein--Sehgal model.

We expect the present model to provide accurate coherent pion
production total and differential cross sections in the first
resonance region, where the $\Delta(1232)$ plays a relevant
role. This energy region is very important for the analysis of present
and forthcoming neutrino oscillation experiments for which good and
reliable theoretical calculations are needed.

\begin{acknowledgments}
  We warmly thank L. Alvarez-Ruso,  G. T. Garvey, F. Sanchez, M. Sorel, M.J.
  Vicente-Vacas and G. Zeller for useful discussions. This research
  was supported by DGI and FEDER funds, under contracts FIS2005-00810,
  FIS2006-03438, FPA2007-65748, and the Spanish Consolider-Ingenio
  2010 Programme CPAN (CSD2007-00042), by Junta de Andaluc\'\i a and
  Junta de Castilla y Le\'on under contracts FQM0225 and SA016A07, and
  it is part of the EU integrated infrastructure initiative Hadron
  Physics Project under contract number RII3-CT-2004-506078.
\end{acknowledgments}



\begin{thebibliography}{blabla}

\bibitem{AguilarArevalo:2007it}
  A.~A.~Aguilar-Arevalo {\it et al.}  [The MiniBooNE Collaboration],
  Phys.\ Rev.\ Lett.\  {\bf 98}, 231801 (2007).

\bibitem{Hiraide:2006zq}
  K.~Hiraide  [SciBooNE Collaboration],
  Nucl.\ Phys.\ Proc.\ Suppl.\  {\bf 159}, 85 (2006).

\bibitem{Marage:1986cy}
  P.~Marage {\it et al.}  [BEBC WA59 COLLABORATION Collaboration],
  Z.\ Phys.\  C {\bf 31}, 191 (1986).

\bibitem{Grabosch:1985mt}
  H.~J.~Grabosch {\it et al.}  [SKAT Collaboration],
  Z.\ Phys.\  C {\bf 31}, 203 (1986).

\bibitem{Allport:1988cq}
  P.~P.~Allport {\it et al.}  [BEBC WA59 Collaboration],
  Z.\ Phys.\  C {\bf 43}, 523 (1989).

\bibitem{Aderholz:1988cs}
  M.~Aderholz {\it et al.}  [E632 Collaboration],
  Phys.\ Rev.\ Lett.\  {\bf 63}, 2349 (1989).

\bibitem{Vilain:1993sf}
  P.~Vilain {\it et al.}  [CHARM-II Collaboration],
  Phys.\ Lett.\  B {\bf 313}, 267 (1993).

\bibitem{Willocq:1992fv}
  S.~Willocq {\it et al.}  [E632 Collaboration],
  Phys.\ Rev.\  D {\bf 47}, 2661 (1993).

\bibitem{Rein:1982pf}
  D.~Rein and L.~M.~Sehgal,
  Nucl.\ Phys.\  B {\bf 223}, 29 (1983).

\bibitem{Adler:1964yx}
  S.~L.~Adler,
  Phys.\ Rev.\  {\bf 135}, B963 (1964).

\bibitem{Hasegawa:2005td}
  M.~Hasegawa {\it et al.}  [K2K Collaboration],
  Phys.\ Rev.\ Lett.\  {\bf 95}, 252301 (2005).

\bibitem{Rein:2006di}
  D.~Rein and L.~M.~Sehgal,
  Phys.\ Lett.\  B {\bf 657}, 207 (2007).

\bibitem{Adler:2005ada}
  S.~L.~Adler, {\it arXiV: hep-ph/0505177}.


\bibitem{Berger:2007rq}
C.~Berger and L.~M.~Sehgal,
Phys.\ Rev.\  D {\bf 76}, 113004 (2007); 
erratum Ibid. {\bf 77}, 059901(E) (2008).

\bibitem{Hi08} K. Hiraide {\it et al.} [SciBooNE Collaboration], 
 {\it arXiv:0811.0369 [hep-ex]}.

\bibitem{Faissner:1983ng}
  H.~Faissner {\it et al.},
  Phys.\ Lett.\  B {\bf 125}, 230 (1983).

\bibitem{Isiksal:1984vh}
  E.~Isiksal, D.~Rein and J.~G.~Morfin,
  Phys.\ Rev.\ Lett.\  {\bf 52}, 1096 (1984).

\bibitem{AguilarArevalo:2008xs}
  A.~A.~Aguilar-Arevalo {\it et al.}  [MiniBooNE Collaboration],
  Phys.\ Lett.\  B {\bf 664}, 41 (2008).

\bibitem{Nachtmann:1970yv}
  O.~Nachtmann,
  Nucl.\ Phys.\  B {\bf 22}, 385 (1970).

\bibitem{Bell:1964eu}
  J.~S.~Bell,
  Phys.\ Rev.\ Lett.\  {\bf 13}, 57 (1964).

\bibitem{Pais:1974kd}
  A.~Pais and S.~B.~Treiman,
  Phys.\ Rev.\  D {\bf 9}, 1459 (1974).

\bibitem{Lackner:1979ax}
  K.~S.~Lackner,
  Nucl.\ Phys.\  B {\bf 153}, 526 (1979).

\bibitem{Belkov:1986hn}
  A.~A.~Belkov and B.~Z.~Kopeliovich,
  Sov.\ J.\ Nucl.\ Phys.\  {\bf 46}, 499 (1987)
  [Yad.\ Fiz.\  {\bf 46}, 874 (1987)].

\bibitem{Kopeliovich:2004px}
  B.~Z.~Kopeliovich,
  Nucl.\ Phys.\ Proc.\ Suppl.\  {\bf 139}, 219 (2005).

\bibitem{Gershtein:1980vd}
  S.~S.~Gershtein, Yu.~Y.~Komachenko and M.~Y.~Khlopov,
  Sov.\ J.\ Nucl.\ Phys.\  {\bf 32}, 861 (1980).

\bibitem{Komachenko:1983jv}
  Yu.~Y.~Komachenko and M.~Y.~Khlopov,
  Yad.\ Fiz.\  {\bf 45}, 467 (1987).

\bibitem{Paschos:2005km}
  E.~A.~Paschos, A.~Kartavtsev and G.~J.~Gounaris,
  Phys.\ Rev.\  D {\bf 74}, 054007 (2006).

\bibitem{Kim:1996az}
  H.~C.~Kim, S.~Schramm and C.~J.~Horowitz,
  Phys.\ Rev.\  C {\bf 53}, 3131 (1996).


\bibitem{Kelkar:1996iv}
  N.~G.~Kelkar, E.~Oset and P.~Fernandez de Cordoba,
  Phys.\ Rev.\  C {\bf 55}, 1964 (1997).

\bibitem{Singh:2006bm}
  S.~K.~Singh, M.~Sajjad Athar and S.~Ahmad,
  Phys.\ Rev.\ Lett.\  {\bf 96}, 241801 (2006).


\bibitem{AlvarezRuso:2007tt}
  L.~Alvarez-Ruso, L.~S.~Geng, S.~Hirenzaki and M.~J.~Vicente Vacas,
  Phys.\ Rev.\  C {\bf 75}, 055501 (2007).
  

\bibitem{AlvarezRuso:2007it}
  L.~Alvarez-Ruso, L.~S.~Geng and M.~J.~Vicente Vacas,
  Phys.\ Rev.\  C {\bf 76}, 068501 (2007).

\bibitem{Hernandez:2007qq}
  E.~Hernandez, J.~Nieves and M.~Valverde,
  Phys.\ Rev.\ D {\bf 76}  033005 (2007).

\bibitem{Gil:1997bm}
  A.~Gil, J.~Nieves and E.~Oset,
  Nucl.\ Phys.\  A {\bf 627}, 543 (1997).

\bibitem{Adler:1968tw}
  S.~L.~Adler,
  Annals Phys.\  {\bf 50}  189 (1968).

\bibitem{Llewellyn Smith:1971zm}
  C.~H.~Llewellyn Smith,
  Phys.\ Rept.\  {\bf 3}, 261 (1972).

\bibitem{Schreiner:1973ka}
  P.~A.~Schreiner and F.~Von Hippel,
  Phys.\ Rev.\ Lett.\  {\bf 30}  339 (1973).

\bibitem{AlvarezRuso:1997jr}
  L.~Alvarez-Ruso, S.~K.~Singh and M.~J.~Vicente Vacas,
  Phys.\ Rev.\  C {\bf 57}  2693  (1998).

\bibitem{AlvarezRuso:1998hi}
  L.~Alvarez-Ruso, S.~K.~Singh and M.~J.~Vicente Vacas,
  Phys.\ Rev.\  C {\bf 59}  3386  (1999).

\bibitem{Lalakulich:2005cs}
  O.~Lalakulich and E.~A.~Paschos,
  Phys.\ Rev.\  D {\bf 71}  074003 (2005).

\bibitem{Leitner:2006ww}
  T.~Leitner, L.~Alvarez-Ruso and U.~Mosel,
  Phys.\ Rev.\  C {\bf 73}  065502 (2006).

\bibitem{Paschos:2003qr}
  E.~A.~Paschos, J.~Y.~Yu and M.~Sakuda,
  Phys.\ Rev.\  D {\bf 69}  014013 (2004).

\bibitem{Lalakulich:2006sw}
  O.~Lalakulich, E.~A.~Paschos and G.~Piranishvili,
  Phys.\ Rev.\  D {\bf 74}  014009 (2006).

\bibitem{Fogli:1979cz}
  G.~L.~Fogli and G.~Nardulli,
  Nucl.\ Phys.\  B {\bf 160}, 116 (1979).

\bibitem{Fogli:1979qj}
  G.~L.~Fogli and G.~Nardulli,
  Nucl.\ Phys.\  B {\bf 165}, 162 (1980).

\bibitem{Sato:2003rq}
  T.~Sato, D.~Uno and T.~S.~H.~Lee,
  Phys.\ Rev.\  C {\bf 67}  065201  (2003).

\bibitem{Barish:1978pj}
  S.~J.~Barish {\it et al.},
  Phys.\ Rev.\  D {\bf 19} (1979) 2521.

\bibitem{Radecky:1981fn}
  G.~M.~Radecky {\it et al.},
  Phys.\ Rev.\  D {\bf 25} (1982) 1161.
  [Erratum-ibid.\  D {\bf 26} (1982) 3297].

\bibitem{Alexandrou:2006mc}
  C.~Alexandrou, T.~Leontiou, J.~W.~Negele and A.~Tsapalis,
  Phys.\ Rev.\ Lett.\  {\bf 98}, 052003 (2007).

\bibitem{BarquillaCano:2007yk}
  D.~Barquilla-Cano, A.~J.~Buchmann and E.~Hernandez,
  Phys.\ Rev.\  C {\bf 75}, 065203 (2007).
  [Erratum-ibid.\  C {\bf 77}, 019903 (2008)].

\bibitem{Graczyk:2007bc}
  K.~M.~Graczyk and J.~T.~Sobczyk,
  Phys.\ Rev.\  D {\bf 77}, 053001 (2008).

\bibitem{luis}
L.~Alvarez-Ruso, L.~S.~Geng, M.~J.~Vicente Vacas private
communication. Erratum in preparation.


\bibitem{Kato:2007zzc}
  I.~Kato  [K2K and T2K Collaborations],
  Nucl.\ Phys.\ Proc.\ Suppl.\  {\bf 168}, 199 (2007).



\bibitem{Carrasco:1991we}
 R.~C.~Carrasco, J.~Nieves and E.~Oset,
 Nucl.\ Phys.\  A {\bf 565}, 797 (1993).

\bibitem{Hirenzaki:1993jc}
  S.~Hirenzaki, J.~Nieves, E.~Oset and M.~J.~Vicente-Vacas,
  Phys.\ Lett.\  B {\bf 304}, 198 (1993).


\bibitem{FernandezdeCordoba:1992ky}
  P.~Fernandez de Cordoba, J.~Nieves, E.~Oset and M.~J.~Vicente-Vacas,
  Phys.\ Lett.\  B {\bf 319}  416 (1993).

\bibitem{Boffi:1991nh}
  S.~Boffi, L.~Bracci and P.~Christillin,
  Nuovo Cim.\  A {\bf 104}, 843 (1991).


\bibitem{Peters:1998mb}
  W.~Peters, H.~Lenske and U.~Mosel,
  Nucl.\ Phys.\  A {\bf 640}, 89 (1998).

\bibitem{Drechsel:1999vh}
  D.~Drechsel, L.~Tiator, S.~S.~Kamalov and S.~N.~Yang,
  Nucl.\ Phys.\  A {\bf 660}, 423 (1999).

\bibitem{Hirata:1978wp}
  M.~Hirata, J.~H.~Koch, E.~J.~Moniz and F.~Lenz,
  Annals Phys.\  {\bf 120}, 205 (1979).


\bibitem{Oset:1981ih}
  E.~Oset, H.~Toki and W.~Weise,
  Phys.\ Rept.\  {\bf 83}, 281 (1982).

\bibitem{Oset:1987re}
  E.~Oset and L.~L.~Salcedo,
  Nucl.\ Phys.\  A {\bf 468}, 631 (1987).



\bibitem{Nieves:1993ev}
  J.~Nieves, E.~Oset and C.~Garcia-Recio,
  Nucl.\ Phys.\  A {\bf 554}, 509 (1993).

\bibitem{Nieves:1991ye}
  J.~Nieves, E.~Oset and C.~Garcia-Recio,
  Nucl.\ Phys.\  A {\bf 554}, 554 (1993).

\bibitem{Nieves:2005rq}
  J.~Nieves, M.~Valverde and M.~J.~Vicente Vacas,
  Phys.\ Rev.\  C {\bf 73}, 025504 (2006).


\bibitem{De Jager:1974dg}
  C.~W.~De Jager, H.~De Vries and C.~De Vries,
  Atom.\ Data Nucl.\ Data Tabl.\  {\bf 14}, 479 (1974).

\bibitem{Negele:1975zz}
  J.~W.~Negele and D.~Vautherin,
  Phys.\ Rev.\  C {\bf 11}, 1031 (1975), and references therein.

\bibitem{GarciaRecio:1991wk}
  C.~Garcia-Recio, J.~Nieves and E.~Oset,
  Nucl.\ Phys.\  A {\bf 547}, 473 (1992).

\bibitem{bnl} T. Kitagaki {\it et al.}, Phys. Rev. {\bf D34} 
2554  (1986).

\bibitem{zeller} G. Zeller, private communication.

\bibitem{Link} J. Link talk at {\it 'Fifth International Workshop on
  Neutrino-Nucleus Interaction in the Few GeV Region (Nuint07)', Fermilab,
  Batavia, Illinois, USA.} http://conferences.fnal.gov/nuint07/
\bibitem{Rein:1980wg}
  D.~Rein and L.~M.~Sehgal,
  Annals Phys.\  {\bf 133}, 79 (1981).

\bibitem{Hawker:2005vf}
  E.~A.~Hawker,
  Nucl.\ Phys.\ Proc.\ Suppl.\  {\bf 139}, 260 (2005).


\bibitem{Nieves:1992pm}
  J.~Nieves and E.~Oset,
  Phys.\ Rev.\  C {\bf 47}  1478  (1993).

\bibitem{Straub:1992yw}
  U.~Straub, J.~Nieves, A.~Faessler and E.~Oset,
  Nucl.\ Phys.\  A {\bf 556}  531 (1993).

\bibitem{Albertus:2002kk}
  C.~Albertus, J.~E.~Amaro and J.~Nieves,
  Phys.\ Rev.\  C {\bf 67}  034604 (2003).

\bibitem{Raaf} J.L. Raaf, PhD thesis, University of Cincinnati,
  FERMILAB-THESIS-2007-20 (2005).

\end{thebibliography}
\end{document}